\documentclass[acmsmall]{acmart}
\usepackage{paralist} 
\usepackage{amsfonts}
\usepackage{blindtext}
\usepackage{amssymb}
\usepackage[linesnumbered,lined,boxed,commentsnumbered]{algorithm2e}
\SetAlFnt{\small}

\usepackage{amsmath}
\usepackage{xcolor} 
\usepackage{framed}
\usepackage{subcaption}
\usepackage[flushleft]{threeparttable}
\usepackage{multicol}
\usepackage{makecell}
\usepackage{array}
\usepackage{eqnarray}
\usepackage{mathrsfs}
\usepackage{graphicx}
\usepackage{textcomp}
\usepackage{enumitem}
\usepackage{longtable}
\usepackage{tabu}
\usepackage{multirow}
\usepackage{breqn}  
\usepackage{mathtools}
\usepackage{caption,fixltx2e}
\usepackage{tabularx}
\usepackage{booktabs} 
\usepackage[T1]{fontenc} 
\usepackage{diagbox} 
\newcolumntype{M}[1]{>{\centering\arraybackslash}m{#1}}

\makeatletter
\newcommand{\thickhline}{%
	\noalign {\ifnum 0=`}\fi \hrule height 2pt
	\futurelet \reserved@a \@xhline
}
\newcolumntype{"}{@{\hskip\tabcolsep\vrule width 1pt\hskip\tabcolsep}}
\makeatother

\newcommand{\tool}{{{{CoRe}}}\xspace} 

\AtBeginDocument{%
  \providecommand\BibTeX{{%
    \normalfont B\kern-0.5em{\scshape i\kern-0.25em b}\kern-0.8em\TeX}}}

\begin{document}

\title{Automating App Review Response Generation Based on Contextual Knowledge}

\author{Cuiyun Gao}
\email{gaocuiyun@hit.edu.cn}
\affiliation{%
  \institution{Harbin Institute of Technology (Shenzhen)}
  \country{China}
}

\author{Wenjie Zhou}
\email{zhouwenjie2@stu.swun.edu.cn}
\affiliation{%
  \institution{Southwest Minzu University}
  \country{China}
}


\author{Xin Xia}
\email{xin.xia@monash.edu}
\affiliation{%
  \institution{Monash University}
\country{Australia}
}

\author{David Lo}
\email{davidlo@smu.edu.sg}
\affiliation{%
  \institution{Singapore Management University}
  \country{Singapore}
  }


\author{Qi Xie}
\email{qi.xie.swun@gmail.com}
\authornote{Corresponding author.}
\affiliation{%
  \institution{Southwest Minzu University}
  \country{China}
  }

\author{Michael R. Lyu}
\email{lyu@cse.cuhk.edu.hk}
\affiliation{%
  \institution{The Chinese University of Hong Kong}
  \city{Hong Kong}
  \country{China}
}

\renewcommand{\shortauthors}{Gao. et al.}

\begin{abstract}
User experience of mobile apps is an essential ingredient that can influence the audience volumes and app revenue. To ensure good user experience and assist app development, several prior studies resort to analysis of app reviews, a type of app repository that directly reflects user opinions about the apps. Accurately responding to the app reviews is one of the ways to relieve user concerns and thus improve user experience. However, the response quality of the existing method relies on the pre-extracted features from other tools, including manually-labelled keywords and predicted review sentiment, which may hinder the generalizability and flexibility of the method. In this paper, we propose a novel end-to-end neural network approach, named \tool, with the contextual knowledge naturally incorporated and without involving external tools. Specifically, \tool integrates two types of contextual knowledge in the training corpus, including official app descriptions from app store and responses of the retrieved semantically similar reviews, for enhancing the relevance and accuracy of the generated review responses. Experiments on practical review data show that \tool can outperform the state-of-the-art method by 11.53\% in terms of BLEU-4, an accuracy metric that is widely used to evaluate text generation systems.

\end{abstract}



\maketitle

\section{Introduction}\label{sec:intro}
According to the report released by~\cite{appdownload}, there are over five billion mobile users worldwide, with global internet penetration standing at 57\%. For these app users, they could choose the apps for usage from a vast number of mobile apps, for example, Google Play and Apple's App Store provide 2.5 million and 1.8 million apps, respectively~\cite{appnumber}. An essential factor for apps to be successful is to guarantee the quality of app functionalities and ensure good user experience. User reviews, which serve as a communication channel between users and developers, can reflect immediate user experience, including app bugs and features to add or modify. Recent research has leveraged natural language processing and machine learning techniques to extract useful information from user reviews to help developers realize, test, optimize, maintain and categorize apps (see e.g., \cite{HarmanAJMSZ16,DBLP:conf/sigsoft/SorboPASVCG16,Grano2018,DBLP:conf/icse/GaoZLK18,DBLP:conf/wcre/CiurumeleaSPG17}) for ensuring good user experience. 

The app stores such as Google Play and App Store also allow developers to respond to the reviews~\cite{googleresponse,iosresponse}, and encourage them to conduct review response promptly and precisely for creating a better user experience and improving app ratings. A recent study by Hassan et al.~\cite{DBLP:journals/ese/HassanTBH18} confirmed the positive effects of review reply. Specifically, they found that responding to a review increases the chances of a user updating their given rating by up to six times in comparison with no responding. McIlroy et al.~\cite{DBLP:journals/software/McIlroySAH17} discovered that users change their ratings 38.7\% of the time following a developer response, with a median increase of 20\% in the rating. Despite of the advantage of review response, developers of many apps never respond to the reviews~\cite{DBLP:journals/ese/HassanTBH18,DBLP:journals/software/McIlroySAH17}. One major reason is the plentiful reviews received for the mobile apps, e.g., the Facebook app on Google Play collects thousands of reviews per day~\cite{apprevenue}. It is labor-intensive and time-consuming for developers to respond to each piece of review. Therefore, the prior work~\cite{DBLP:conf/kbse/GaoZX0LK19} initiates automating the review response process.

\begin{figure}[ht]
    \centering
    \begin{subfigure}[b]{0.6\textwidth}
        \includegraphics[width=\textwidth]{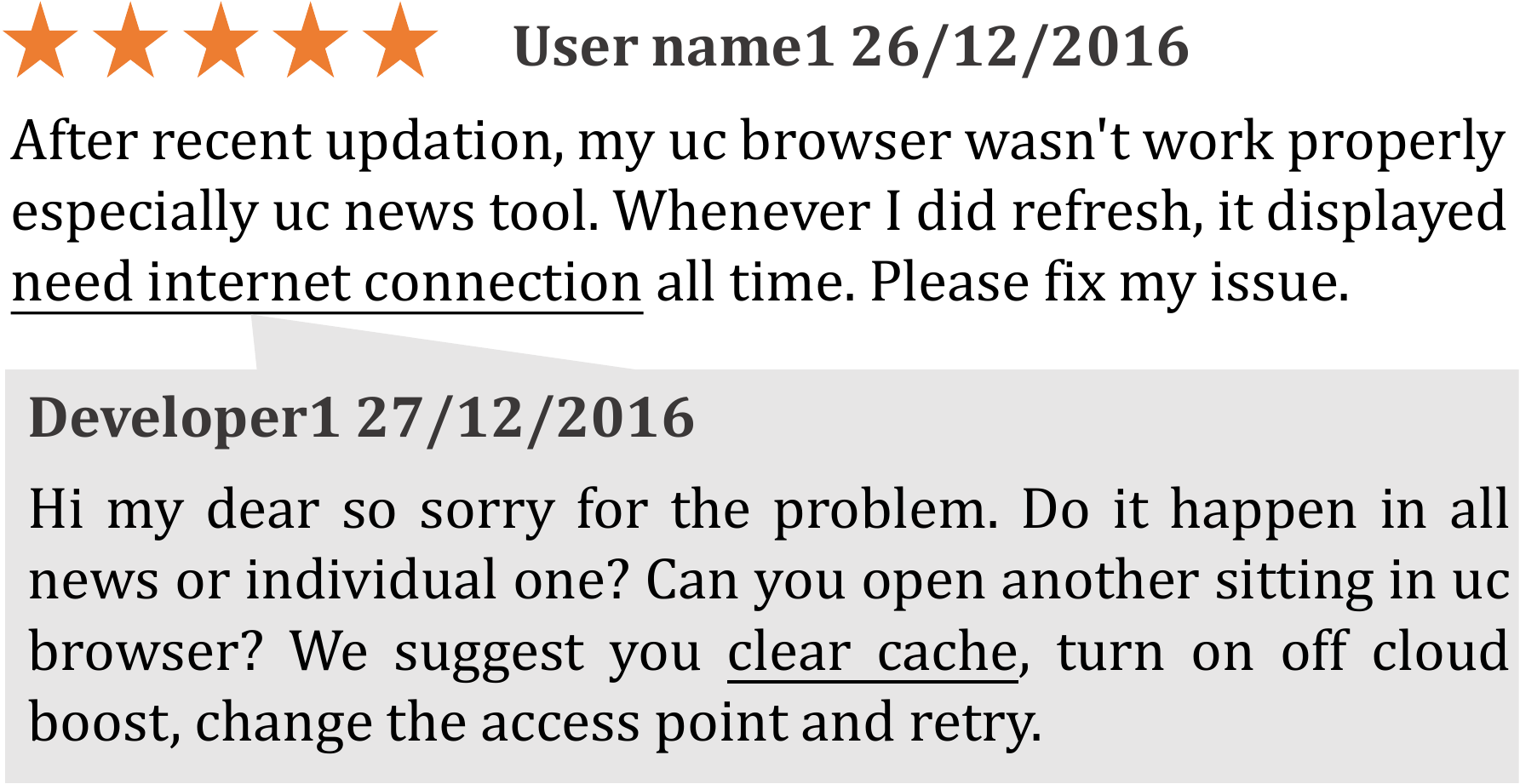}
        \caption{One review instance of the UC Browser app.}
        \label{fig:example1}
      \end{subfigure}
      \vfill
      \begin{subfigure}[b]{0.62\textwidth}
        \includegraphics[width=\textwidth]{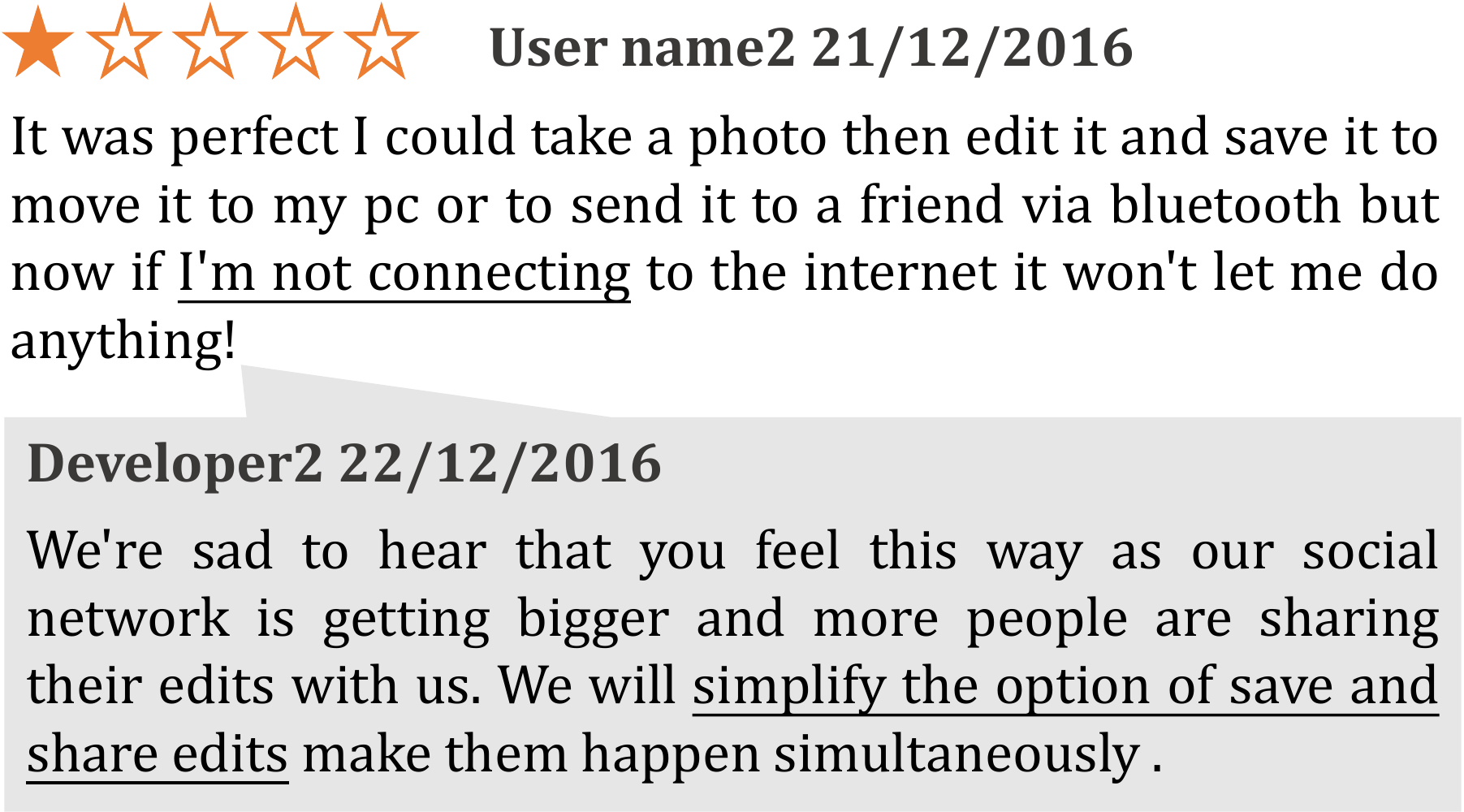}
        \caption{One review instance of the PicsArt Photo Editor app.}
        \label{fig:example2}
      \end{subfigure}
    \caption{Review instances from two separate apps. The underlined texts highlight the main issues reported in reviews and corresponding suggested solutions from developers.}
	\label{fig:examples}
\end{figure}

Review response generation can be analogical to social dialogue generation~\cite{DBLP:conf/acl/WuLZZW20,DBLP:conf/emnlp/LiMRJGG16} in the natural language processing field. Different from social dialogue generation, app review-response generation is more domain-specific or even app-specific, and hence, its performance strongly relies on the establishment of the domain knowledge. For example, the response for the review of one app may not be applicable for the review of another app even though the reflected issues are similar. As illustrated in Figure~\ref{fig:examples}, both review instances are complaining about the Internet connection issue, but developers' suggested solutions are different. For the UC browser app, the developer suggests to clear cache while for the PicsArt photo editor app, the developers undertake to simplify the options of save and share edits.

To automatically learn the domain-specific knowledge, Gao et al.~\cite{DBLP:conf/kbse/GaoZX0LK19} proposed a Neural Machine Translation (NMT)~\cite{DBLP:conf/nips/SutskeverVL14}-based neural network, named RRGen, which can encode user reviews with an embedding layer and decode them into developers' response through a Gated Recurrent Unit (GRU)~\cite{DBLP:conf/emnlp/ChoMGBBSB14} model with attention mechanism. External review attributes including review length, rating, predicted sentiment, app category, and pre-defined keywords, are adopted to better encode the semantics of user reviews. Although good performance is demonstrated, the design of RRGen exhibits two main limitations. First, RRGen highly relies on the performance of the external tools such as SURF~\cite{DBLP:conf/sigsoft/SorboPASVCG16} for determining pre-defined keywords and SentiStrength~\cite{DBLP:journals/jasis/ThelwallBPCK10} for estimating review sentiment. This weakens the flexibility and generalization of RRGen, e.g., when keywords in the reviews are not in the pre-defined keyword dictionary. Second, RRGen presents the similar problem of NMT-based approaches, i.e., they generally prefer high-frequency words in the corpus and the generated responses are often generic and not informative~\cite{DBLP:conf/emnlp/ArthurNN16,DBLP:conf/naacl/ZhangUSNN18,DBLP:conf/cikm/0005HQQGCLSL19}.



To alleviate the above limitations, we propose a novel neural architecture namely \textbf{C}ontextual kn\textbf{o}wledge-based app \textbf{R}eview r\textbf{e}sponse generation (\tool), built upon official app descriptions and responses of retrieved similar reviews from the training corpus. For mitigating the first limitation, we incorporate app descriptions, which usually contain sketches of app functionalities~\cite{DBLP:conf/esem/Al-SubaihinSBCH16}. Based on app descriptions, the neural model can learn to pay attention to app functionality-related words in the reviews, without feeding pre-defined keywords into the model. For relieving the second limitation, we involve responses of similar reviews based on Information Retrieval (IR)-based approach. The IR-based approach~\cite{DBLP:journals/corr/JiLL14} has proven useful in leveraging the responses of similar conversations for producing relevant responses, so the IR-based retrieved responses are highly probable to contain the words in the expected responses (including the low-frequency ones). To incorporate the words in the retrieved responses, \tool utilizes pointer-generator network~\cite{DBLP:conf/acl/SeeLM17} to adaptively copy words from the responses instead of simply from a fixed vocabulary obtained from the training corpus.

Experiments based on 309,246 review-response pairs from 58 popular apps show that \tool significantly outperforms the state-of-the-art model by 11.53\% in terms of BLEU-4 score~\cite{DBLP:conf/acl/PapineniRWZ02} (An accuracy measure that is widely used to evaluate text generation systems). Human study with 20 programmers through Tencent Online Questionnaire~\cite{tencentwj} further confirms that \tool can generate a more relevant and accurate response than RRGen. 

The remainder of this paper is organized as follows. Section~\ref{sec:back} introduces the background of our work. Section~\ref{sec:method} illustrates the proposed approach. Section~\ref{sec:setup} and Section~\ref{sec:result} detail our experimental settings and the experimental results, respectively. Section~\ref{sec:human} describes the human evaluation results.
Section~\ref{sec:discussion} discusses the advantages of the proposed approach and threats to validity. Section~\ref{sec:literature} surveys the related work. Section~\ref{sec:conclusion} concludes the paper.

\section{Background}\label{sec:back}
In this section, we introduce the background knowledge of the proposed approach, including attentional encoder-decoder model and pointer-generator model.

\subsection{Attentional Encoder-Decoder Model}

\begin{figure}[h]
    \centering
    \includegraphics[width=0.65\textwidth]{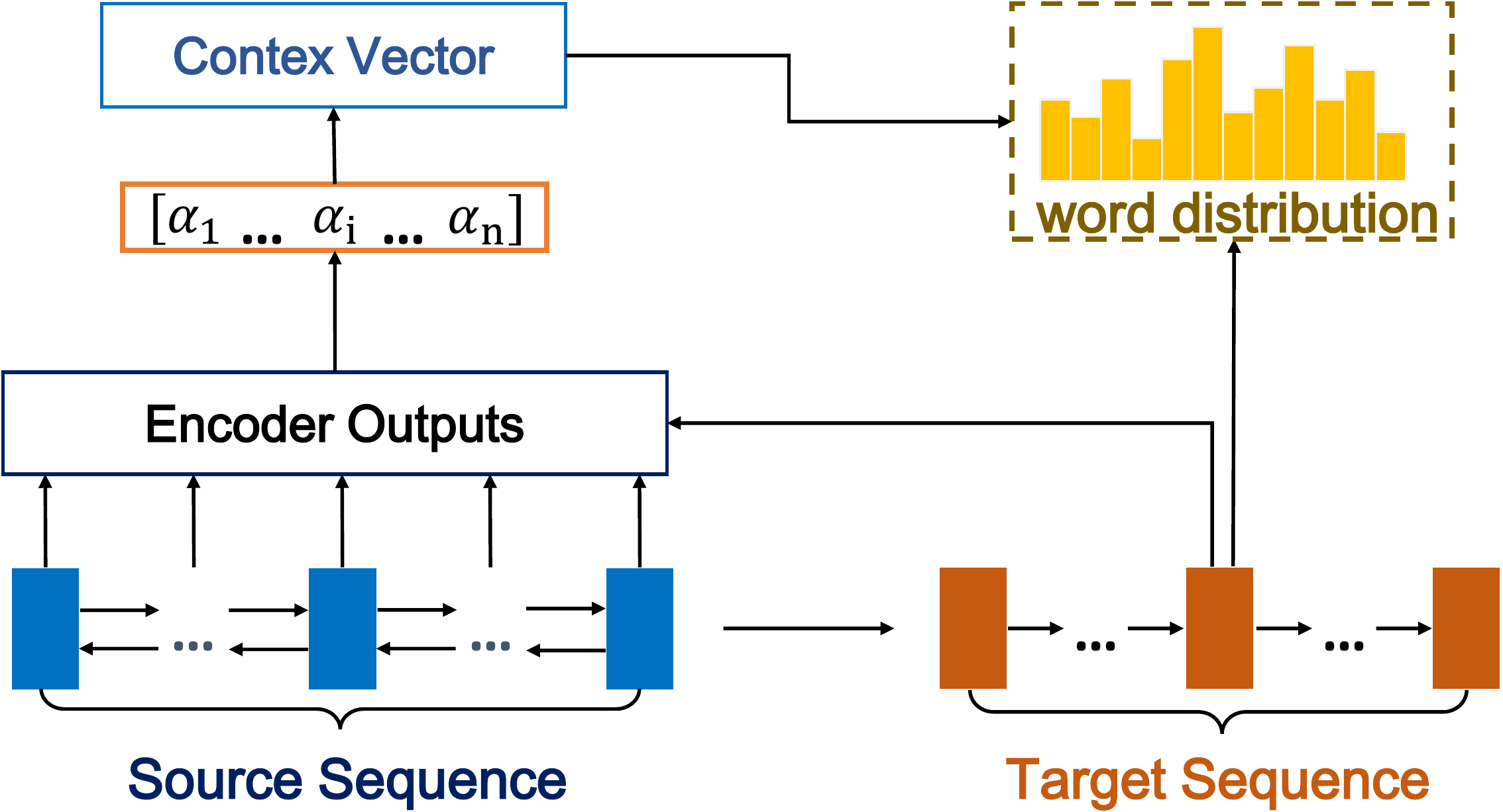}
    \caption{Graphical illustration of the attentional bi-LSTM encoder-decode model.}
    \label{fig:attention}
\end{figure}

Encoder-decoder model, also called sequence-to-sequence model, has demonstrated the ability to model the variable-length input and output, e.g., words and sentences. Figure~\ref{fig:attention} illustrates the architecture of the attentional encoder-decoder model. Generally, tokens of the source sequence $\mathbf{w}=(w_1, w_2, ..., w_n)$ ($n$ is the number of input tokens) are fed one-by-one into the encoder (a single-layer bidirectional GRU~\cite{DBLP:conf/emnlp/ChoMGBBSB14} as shown in Figure~\ref{fig:attention}), producing a sequence of encoder hidden states $\mathbf{h}=(h_1, h_2,...,h_{n})$. On each step $t$, the decoder (a single-layer unidirectional GRU) is often trained to predict the next word $y_t$ based on the context vector $\mathbf{c}$ and previously predicted words $\{y_1, ..., y_{t-1}\}$, and has decoder state $s_t$. The context vector $c_t$ depends on a sequence of encoder hidden states $\mathbf{h}$, and is computed as a weighted sum of the hidden states~\cite{DBLP:journals/corr/BahdanauCB14}:

\begin{equation}\label{equ:attention}
\begin{split}
c_t & = \sum_{j}^n \alpha_{tj}h_j, \\
\alpha_{tj} & = \text{softmax}(e_{tj}),
\end{split}
\end{equation}


\noindent where $e_{tj}$ measures the similarity degree between the input hidden state $h_j$ and decoder state $s_{t-1}$. The attention weight $\alpha_t$ can be viewed as a probability distribution over the source words, and higher probabilities render the decoder pay more attention to the corresponding input during producing the next word. The context vector is then concatenated with the decoder state $s_t$ and fed through two linear layers to generate the vocab distribution:

\begin{equation}\label{equ:vocab}
P_t^{\text{vocab}} (w)=\text{softmax}(v'(v[s_t,c_t]+b)+b'),
\end{equation}

\noindent where $v$, $v'$, $b$, and $b'$ are learnable parameters, and $P_t^{\text{vocab}}$ is a probability distribution over all the words in the vocabulary. The model is trained to minimize the negative log likelihood:

\begin{equation}\label{equ:loss}
    \text{loss} = \text{min}\frac{1}{N}\sum_i-\log P(y_i|x_i),
\end{equation}

\noindent where each $(x_i,y_i)$ is a (source sequence, target sequence) pair from the training set.

\subsection{Pointer-Generator Model}
\begin{figure}[h]
    \centering
    \includegraphics[width=0.75\textwidth]{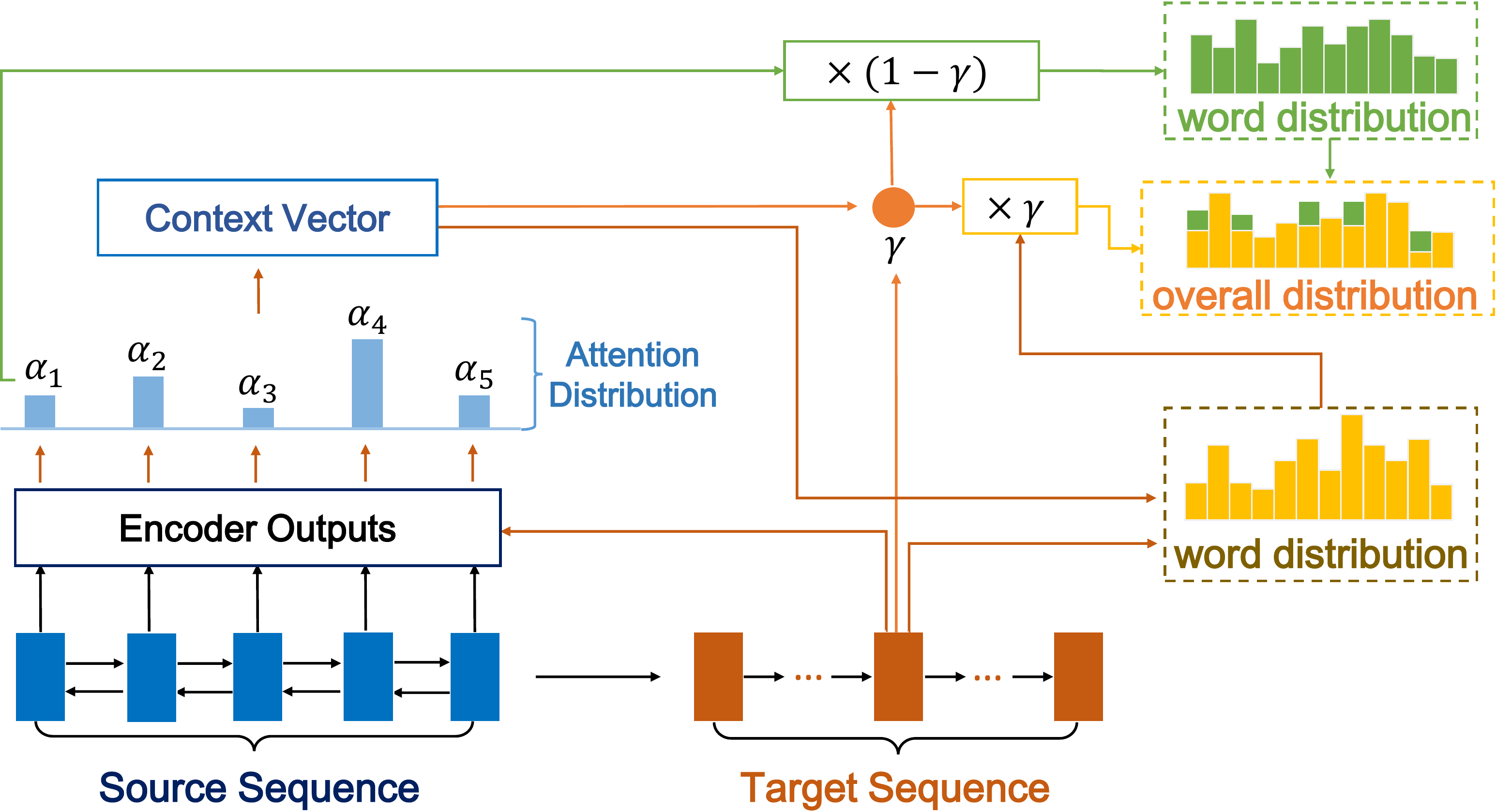}
    \caption{Graphical illustration of the pointer-generator model.}
    \label{fig:pointer}
\end{figure}

Pointer-generator networks~\cite{DBLP:conf/nips/VinyalsFJ15,DBLP:conf/acl/SeeLM17} allow sequence-to-sequence models to predict words during decoding by either copying words via pointing or generating words from a fixed vocabulary. Figure~\ref{fig:pointer} depicts the architecture of the point-generator model. As can be seen, besides computing the context vector $c_t$ and attention weight $\alpha_t$, the generation probability $\gamma_t\in[0,1]$ for step $t$ is calculated for the context vector $c_t$, the decoder state $s_t$ and the decoder input $w_t$:

\begin{equation}
    \gamma_t = \sigma(\omega^\intercal_cc_t+\omega^\intercal_ss_t+\omega^\intercal_ww_t+b_{ptr}),
\end{equation}

\noindent where vectors $\omega_c$, $\omega_s$, $\omega_w$ and scalar $b_{ptr}$ are learnable parameters. $\sigma$ is the sigmoid function. $\gamma_t$ can be regarded as an indicator of which source the predicted word comes from. The probability distribution over the \textit{overall vocabulary} is computed as:

\begin{equation}
    P_t(w) = \gamma_t\cdot P_t^{\text{vocab}}(w)+(1-\gamma_t)\cdot \sum_{i:w_i=w} \alpha_{ti}.
\end{equation}

\noindent If $w$ is an out-of-vocabulary (OOV) word, then $P_t^{\text{vocab}}(w)$ is zero. In this way, point-generator models are able to generate OOV words. The loss function is the same as described in equations (\ref{equ:loss}).
\section{Methodology}\label{sec:method}

\begin{figure}[ht]
    \centering
    \includegraphics[width=\textwidth]{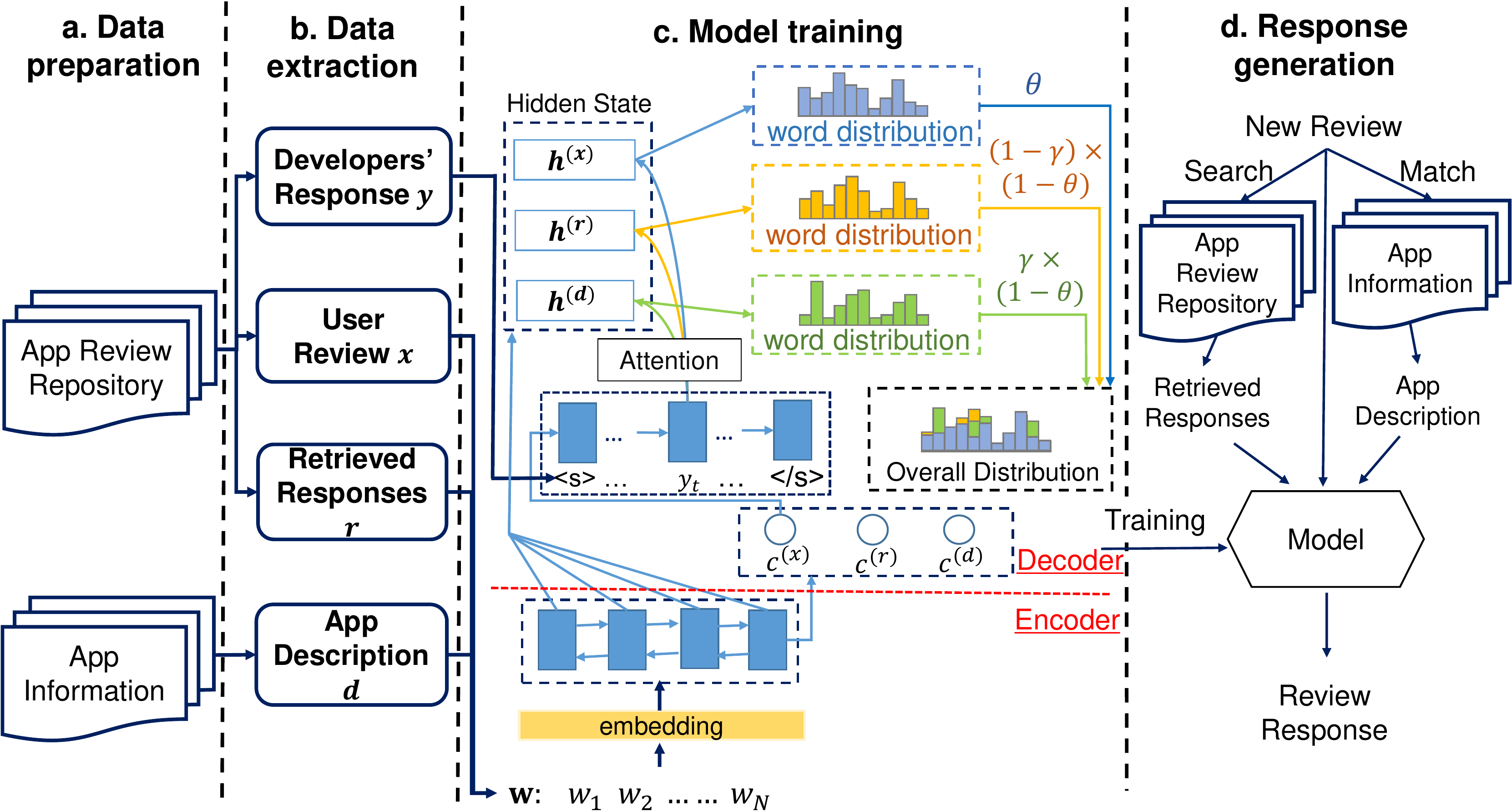}
    \caption{Overall architecture of \tool.}
    \label{fig:framework}
\end{figure}

This section describes our proposed model \tool, which builds upon the basic pointer-generator model. Besides user reviews, two types of contextual knowledge, including app descriptions and responses of the retrieved similar reviews from the training corpus, are regarded as the source sequence. The developers' responses are treated as the target sequence. App descriptions generally describe apps' functionalities~\cite{DBLP:conf/esem/Al-SubaihinSBCH16}, so with app descriptions integrated, the words related to app functionalities are prone to be captured. Semantically-similar reviews are involved since the semantics of the corresponding responses tend to be identical. For each piece of review, the semantic distances with other reviews in the training set are computed as the cosine similarity between the unigram tf-idf representations, and only the responses of the top $K$ reviews with highest similarity scores are considered for the response generation.

The overall architecture of the proposed model is illustrated in Figure~\ref{fig:framework}. \tool is mainly composed of four stages: Data preparation, data extraction, model training, and response generation. We first preprocess the app reviews, their responses and app descriptions collected from Google Play. The processed data are then parsed into a parallel corpus of user reviews, corresponding responses, the retrieved responses, and app description. Based on the parallel corpus, we build and train a pointer-generator-based model with the contextual knowledge holistically considered. The details are elaborated in the following.

\subsection{Source Sequence Encoding}
Let $\mathbf{w}=(w_1, w_2,...,w_n)$ be a sequence of source tokens, which can be the input review $\mathbf{x}$, app description $\mathbf{d}$ or the response for each of top K retrieved similar reviews $\mathbf{r}^{(k)},1 \leq k\leq K$. We first obtain a trainable embedded representation of each token in the sequence and then adopt bi-GRU to encode the sequence of the embedding vectors. 

\begin{equation}
    e^{(x)},\mathbf{h}^{(x)} = \text{bi-GRU}(\mathbf{x}),
\end{equation}
\begin{equation}\label{equ:des}
    e^{(d)},\mathbf{h}^{(d)} = \text{bi-GRU}(\mathbf{d}),
\end{equation}
\begin{equation}
    e^{(r)(k)},\mathbf{h}^{(r)(k)} = \text{bi-GRU}(\mathbf{r}^{(k)}),
\end{equation}

\noindent where $e^\Delta$ and $\mathbf{h}^\Delta=(h_1, h_2,..., h_n)$ denote the final hidden state of the bi-LSTM and outputs of bi-LSTM at all steps, where $\Delta\in [(x), (d), (r)(1),..., (r)(k),..., (r)(K)]$.

\subsection{Contextual Knowledge Integration}\label{sec:fusion}
Different from the basic pointer-generator network~\cite{DBLP:conf/acl/SeeLM17}, \tool also allows integrating tokens from the contextual information besides the input reviews. At decoder step $t$, the decoder state $s_t$ is used to attend over the app description tokens and the retrieved response tokens to produce a probability distribution over the tokens appearing in the description and retrieved responses respectively. These distributions are then integrated with the attention distribution obtained by the decoder over the fixed vocabulary to compute an overall distribution.

\subsubsection{Copying tokens from app description}
Similar to the basic attentional encoder-decoder model, we encode the description tokens $\mathbf{d}$ and apply attention to the encoder outputs at a decoder step $t$. This produces the attention weights $\alpha_t^{(d)}$ and a representation of the entire context $c_t^{(d)}$. The context vector is then employed to obtain the probability distribution $P_t^{(d)}(w)$ over the tokens in the app description:

\begin{equation}\label{equ:att_desc}
    \alpha_t^{(d)}, c_t^{(d)} = \text{Attention}(\mathbf{h}^{(d)}, s_t),
\end{equation}
\begin{equation}
    P_{t}^{(d)}(w) = g(s_t,y_{t-1}, c_{t}^{(d)}),
\end{equation}

\noindent where $\mathbf{h}^{(d)}$ indicates the encoder outputs as computed in Equation (\ref{equ:des}) and $g$ is a non-linear mapping function.

\subsubsection{Copying tokens from responses of the retrieved reviews}
\begin{figure}[ht]
    \centering
    \includegraphics[width=\textwidth]{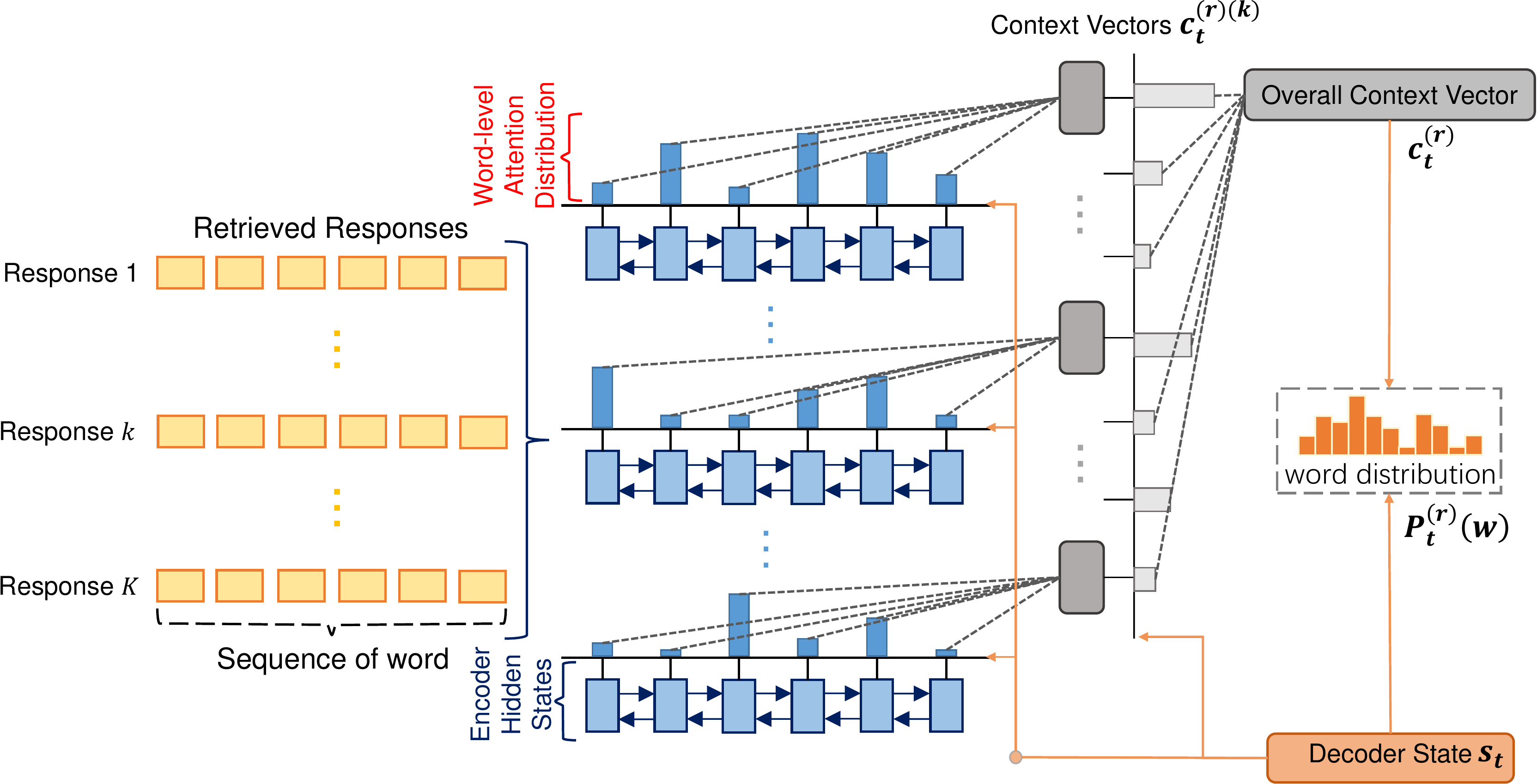}
    \caption{Illustration of the hierarchical pointer network for copying tokens from the retrieved $K$ responses.}
    \label{fig:copy_desc}
\end{figure}

To integrate the responses of the $K$ retrieved reviews, we adapt the hierarchical pointer network as shown in Figure~\ref{fig:copy_desc} for involving tokens from multiple extracted responses. Based on the token-level representations $\mathbf{h}^{(r)(k)}$, the decoder state $s_t$ is used to attend over the tokens in each retrieved response:

\begin{equation}
    \alpha_t^{(r)(k)}, c_t^{(r)(k)} = \text{Attention}(\mathbf{h}^{(r)(k)}, s_t),
\end{equation}
\begin{equation}\label{equ:re-attention}
    \alpha_t^{(r)},c_t^{(r)} = \text{Attention}([c_t^{(r)(1)},... ,c_t^{(r)(K)}], s_t),
\end{equation}
\begin{equation}
     P_t^{(r)}(w) = g(s_t,y_{t-1}, c_t^{(r)}),
\end{equation}

\noindent where $\mathbf{h}^{(r)(k)}$ is the output of the encoder for the response of the top $k$-th retrieved reviews. The context vector $c_t^{(r)}$ for all the retrieved responses are obtained based on the context vectors of all the $K$ responses, following the Equation (\ref{equ:re-attention}). $P_t^{(r)}(w)$ means the probability distribution over the tokens in the retrieved $K$ responses. 

\subsubsection{Attention fusion}
We first fuse the two vocabulary distributions $P_t^{(d)}(w)$ and $P_t^{(r)}(w)$ which represent the probabilities of copying tokens from the app description and retrieved responses respectively. We compute the fused attention vector using the decoder state $s_t$, the overall app description representation $c_t^{(d)}$ and overall retrieved response representation $c_t^{(r)}$ (Equation \ref{equ:fuse}). The computed attention weight $\gamma_t$ is adopted to combine the two copying distributions as Equation (\ref{equ:fuse_p}).

\begin{equation}\label{equ:fuse}
    \gamma_t, c_t^{\text{fuse}} = \text{Attention}([c_t^{(d)},c_t^{(r)}], s_t),
\end{equation}
\begin{equation}\label{equ:fuse_p}
    P_t^{\text{fuse}}(w) = \gamma_t\cdot P_t^{(d)}(w)+(1-\gamma_t)\cdot P_t^{(r)}(w).
\end{equation}

The overall distribution $P_t(w)$ for the training vocabulary at each decoder step $t$ is calculated based on the context vector $c_t^{\text{fuse}}$ of the two contextual sources and decoder state $s_t$.

\begin{equation}
\begin{split}
    \theta_t & = \sigma(\omega_{f}^\intercal c_t^{\text{fuse}}+\omega_s^\intercal s_t+\omega_x^\intercal x_t+b_{\text{ptr}}), \\
    P_t(w) &= \theta_t\cdot P_t^{\text{vocab}}(w)+(1-\theta_t)\cdot P_t^{\text{fuse}}(w),
\end{split}
\end{equation}

\noindent where $\omega_f$, $\omega_s$, $\omega_x$ and $b_{\text{ptr}}$ are learnable parameters, $x_t$ is the decoder input, and $P_t^{\text{vocab}}(w)$ indicates the vocabulary distribution based on the input reviews only (referring to Equation~\ref{equ:vocab}).

\subsection{Model Training and Validation}
\subsubsection{Training} We train the whole network end-to-end with the negative log-likelihood loss function of 

\begin{equation}
    J_{\text{loss}}(\Theta)=-\frac{1}{|y|}\sum_{t=1}^{|y|}\log(p_t(y_t|y<t,\mathbf{x},\mathbf{d},\{\mathbf{r}^{(k)}\}_{k=1}^K)),
\end{equation}

\noindent for a training sample $(\mathbf{x},\mathbf{y},\mathbf{d},\{\mathbf{r}^{(i)}\}_{i=1}^K))$ where $\Theta$ denotes all the learnable model parameters. The attentional encoder-decoder model has various implementations. We adopt bidirectional Gated Recurrent Units (GRUs)~\cite{DBLP:conf/emnlp/ChoMGBBSB14} which is a popular basic encoder-decoder model and performs well in many text generation tasks~\cite{DBLP:journals/corr/ChungGCB14,DBLP:conf/icassp/WuK16}. The hidden units of GRUs are set as 200 and word embeddings are initiated with pre-trained 100-dimensional GloVe vectors~\cite{glove}. The maximum sequence lengths for reviews, app descriptions, and retrieved responses are all defined as 200. We save the model every 200 batches. The number of retrieved responses, the dropout rate, and the number of hidden layers are defined as 4, 0.1, and 1, respectively. Details of parameter tuning are discussed at Section~\ref{sec:parameter}. The whole model is trained using the minibatch Adam~\cite{DBLP:journals/corr/KingmaB14}, a stochastic optimization approach which can automatically adjust the learning rate. The batch size is set as 32. During training the neural networks, we limit the source and target vocabulary to the top 10,000 words that are most frequently appeared in the training set.

For implementation, we use PyTorch~\cite{pytorch}, an open-source deep learning framework. We train our model in a server with Intel(R) Xeon(R) Gold 6230 CPU @ 2.10GHz, Tesla T4 16G. The training lasts $\sim$8 hours with three epochs.

\subsubsection{Validation} We evaluate on the test set after the batch during which the trained model shows an improved performance on the validation set regarding BLEU score~\cite{DBLP:conf/acl/PapineniRWZ02}. The evaluation results are the highest test score and corresponding generated response. We use the same GPU as used in training and the testing process cost around 30 minutes.

\section{Experimental Setup}\label{sec:setup}
In this section, we elaborate on the setup of our experiments, including experimental dataset, the evaluation metric, and baseline approaches.

\subsection{Experimental Dataset}
We perform experiments for verifying the effectiveness of the proposed model on the recently released review response dataset~\cite{DBLP:conf/kbse/GaoZX0LK19}. The dataset includes 309,246 review-response pairs from 58 popular apps, with 279,792, 14,727, and 14,727 pairs in the training, validation, and test sets, respectively. Besides the review-response pairs, we crawled the corresponding app descriptions from Google Play for the 58 subject apps. For the app descriptions, we remove all special characters such as ``$\bigstar$'' and conduct similar preprocessing steps as the review preprocessing steps~\cite{DBLP:conf/kbse/GaoZX0LK19}, including lowercase and lemmatization. After the basic preprocessing, we observe that the maximum, median and minimum lengths of the app descriptions are 625, 300 and 43 words, respectively, with the average length at 314. Since the semantics of long input texts are difficult to be effectively learnt by the basic attentional encoder-decoder model~\cite{DBLP:journals/neco/YuSHZ19}, we reduce the input description lengths by manually filtering out the sentences irrelevant to the app features/functionalities (e.g., the sentences explicitly encouraging users to download the apps, ``\textit{download the highest rated travel app now and join thousands of bookers like you finding unmissable hotel deals!}''). The pruning process costs us around 1.5 hours for the 58 subject apps. The maximum, median and minimum lengths of the reduced descriptions are 198, 151 and 43 words, respectively, with the average length at 146.

\subsection{Evaluation Metric}
BLEU is a metric widely used in natural language processing and software engineering fields to evaluate generative tasks (e.g., machine translation, dialogue generation and code commit message generation)~\cite{DBLP:conf/emnlp/LiMRJGG16,DBLP:conf/naacl/ZhangUSNN18,DBLP:conf/kbse/JiangAM17,DBLP:conf/iwpc/HuLXLJ18}. It calculates the frequencies of the co-occurrence of n-grams between the ground truth $\hat{y}$ and the generated sequence $y$ to judge their similarity. 

\begin{equation}\label{equ:p}
    p_n(y,\hat{y}) = \frac{\sum_j \min(h(j,\hat{y}), h(j,y))}{\sum_j h(j,\hat{y})},
\end{equation}

\noindent where $j$ indexes all possible n-grams, and $h(j,y)$ or $h(j, \hat{y})$ indicate the number of $j$-th n-grams in the generated sequence $y$ or the ground truth $\hat{y}$ respectively. To avoid the drawbacks of using a precision score, namely it favours shorter generated sentences, BLEU-N introduces a brevity penalty.

\begin{equation}
\text{BLEU-N}:=b(y,\hat{y})\exp(\sum_{n=1}^N \beta_n \log p_n(y,\hat{y})),
\end{equation}

\noindent where $b(y,\hat{y})$ is the brevity penalty and $\beta_n$ is a weighting parameter. We use corpus-level BLEU-4, i.e., $N=4$, as our evaluation metric since it is demonstrated to be more correlated with human judgements than other evaluation metrics~\cite{DBLP:conf/emnlp/LiuLSNCP16}.


\subsection{Baseline Approaches}
We compare the performance of the proposed \tool with a random selection approach, the basic attentional encoder-decoder model~\cite{DBLP:journals/corr/BahdanauCB14}, and the state-of-the-art approach for review response generation~\cite{DBLP:conf/kbse/GaoZX0LK19}, namely RRGen. We elaborate on the first and last baselines as below.

\textbf{Random Selection:} The approach randomly picks one response in the training set as the response to a review in the test set.

\textbf{RRGen:} It is the state-of-the-art approach for automating review reply generation. RRGen explicitly combines review attributes, such as review length, rating, predicted sentiment and app category, and occurrences of specific keywords into the basic attentional encoder-decoder (NMT) model.
\section{Experimental Results}\label{sec:result}
In this section, we elaborate on the results of the evaluation of \tool through experiments and compare it with the state-of-the-art tool, RRGen~\cite{DBLP:conf/kbse/GaoZX0LK19}, and another competing approach, NMT~\cite{DBLP:journals/corr/BahdanauCB14}, to assess its capability in accurately responding to user reviews. Our experiments are aimed at answering the following research questions.

\begin{enumerate}[label=\bfseries RQ\arabic*:,leftmargin=.5in]
    \item What is the performance of \tool in responding to user reviews?
    
    \item What is the impact of the involved contextual knowledge on the performance of \tool? 

    
    \item How accurate is \tool under different parameter settings?

\end{enumerate}

\subsection{RQ1: What is the performance of \tool in responding to user reviews?}
Table~\ref{tab:comp_res} illustrate the comparison results with the baseline approaches. As can be seen, the proposed \tool shows the best performance among all the approaches. Specifically, \tool outperforms the three baselines by 11.53\%$\sim$5.16 times. From the $p_n$ scores, we can observe that the responses produced by \tool consist of more similar n-grams comparing to the ground truth. For example, \tool increases the performance of the baselines by at least 13.55\% regarding the accuracy of 4-gram prediction.

We then use Wilcoxon signed-rank test~\cite{wilcoxon1992individual} to verify whether the increase is significant, and Cliff'd Delta (or $d$) to measure the effect size~\cite{DBLP:journals/technometrics/Ahmed06}. The significance test result ($p$-value <0.01) and large effect size on the metrics ($|d|$>0.474) of \tool and RRGen indicate that the proposed model can generate more accurate and relevant responses to user reviews.

\begin{table}[t]
\centering
\caption{Comparison results with baseline approaches. \textbf{Bold} figures highlight better results. $p_n$ indicates the $n$-gram precision computed in Equation (\ref{equ:p}). Statistical significance results are indicated with *($p$-value<0.01).}\label{tab:comp_res}
    \scalebox{1.0}{
        \begin{tabular}{c|r|r|r|r|r}
        \hline
        \hline
           Model  & BLEU-4 & $p_1$ & $p_2$ & $p_3$ & $p_4$\\
        \hline
            Random & 6.55* & 27.64* & 6.90* & 3.55* & 2.78* \\
            NMT &  21.61* & 40.55* & 20.75* &  16.78* & 15.47* \\
            RRGen & 36.17* & 53.24* & 35.83* & 31.73* & 30.04* \\
        \hline
            \tool & \textbf{40.34} & \textbf{57.17} & \textbf{40.24} & \textbf{35.96} & \textbf{34.11} \\
        \hline
        \hline
        \end{tabular}
    }
\end{table}

\subsection{RQ2: What is the impact of the involved contextual knowledge on the performance of \tool?}
We analyze the impact of the involved contextual knowledge, including app description and the retrieved responses, on the model performance. We perform contrastive experiments in which only a single source of contextual information is considered in the basic attentional encoder-decoder model. Table~\ref{tab:extension} illustrates the results.

The integration of both app description and the retrieved responses presents the highest improvements. With either type of contextual information individually combined, the model achieves comparative performance, i.e., $\sim$38 and $\sim$54 in terms of BLEU-4 and $p_1$ scores respectively. However, without the contextual information included, the performance shows dramatic decline, presenting only 20.1 in terms of the BLEU-4 metric. This implies the importance of integrating contextual knowledge for accurate review response generation, and each type of the considered contextual knowledge is helpful for improving the generation accuracy. We analyze deeper into the advantage carried by the contextual knowledge in Section \ref{sec:advantage}.

\begin{table}[t]
\centering
\caption{Contrastive experiments with individual extension removed.}\label{tab:extension}
    \scalebox{1.0}{
        \begin{tabular}{l|r|r|r|r|r}
        \hline
        \hline
           Model  & BLEU-4 & $p_1$ & $p_2$ & $p_3$ & $p_4$\\
        \hline
            \tool & \textbf{40.34} & \textbf{57.17} & \textbf{40.24} & \textbf{35.96} & \textbf{34.11}  \\
        \hline
            -Retrieval & 38.65  & 54.73 & 37.91 & 33.71 & 31.90 \\
            -Description & 38.58 & 54.00 & 36.71 & 32.55 & 30.71  \\
        \hline
        Only review (NMT) & 21.61 & 40.55 & 20.75 &  16.78 & 15.47\\
        \hline
        \hline
        \end{tabular}
    }
\end{table}



\subsection{RQ3: How accurate is \tool under different parameter settings?}\label{sec:parameter}
\begin{figure}[t]
     \centering
     \begin{subfigure}[h]{0.45\textwidth}
        \centering
    	\includegraphics[width=1 \textwidth]{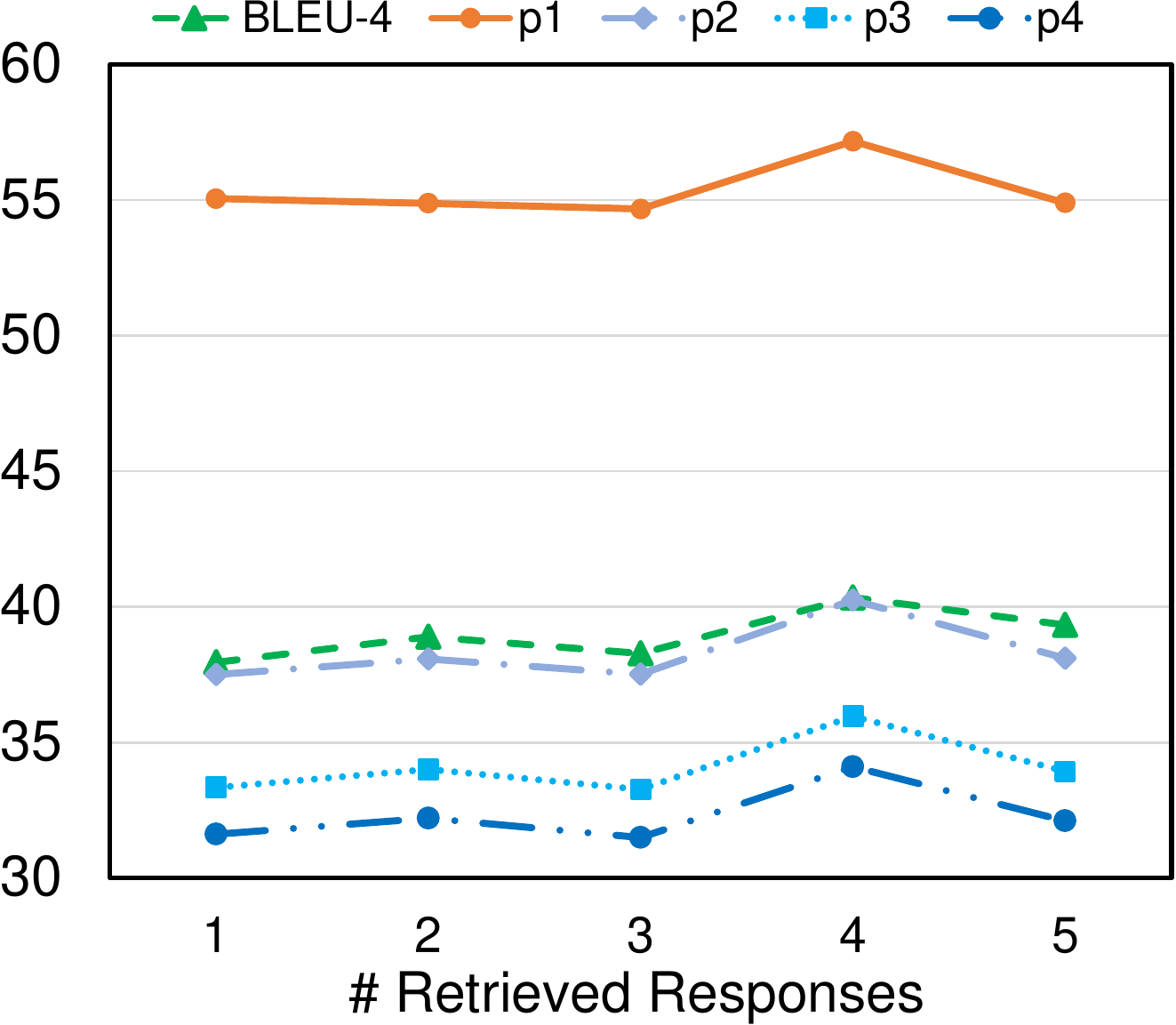}
    	\caption{\small Different numbers of retrieved responses.}
        \end{subfigure}
        \hfill
        \begin{subfigure}[h]{0.45\textwidth}
        \centering
        \includegraphics[width=1 \textwidth]{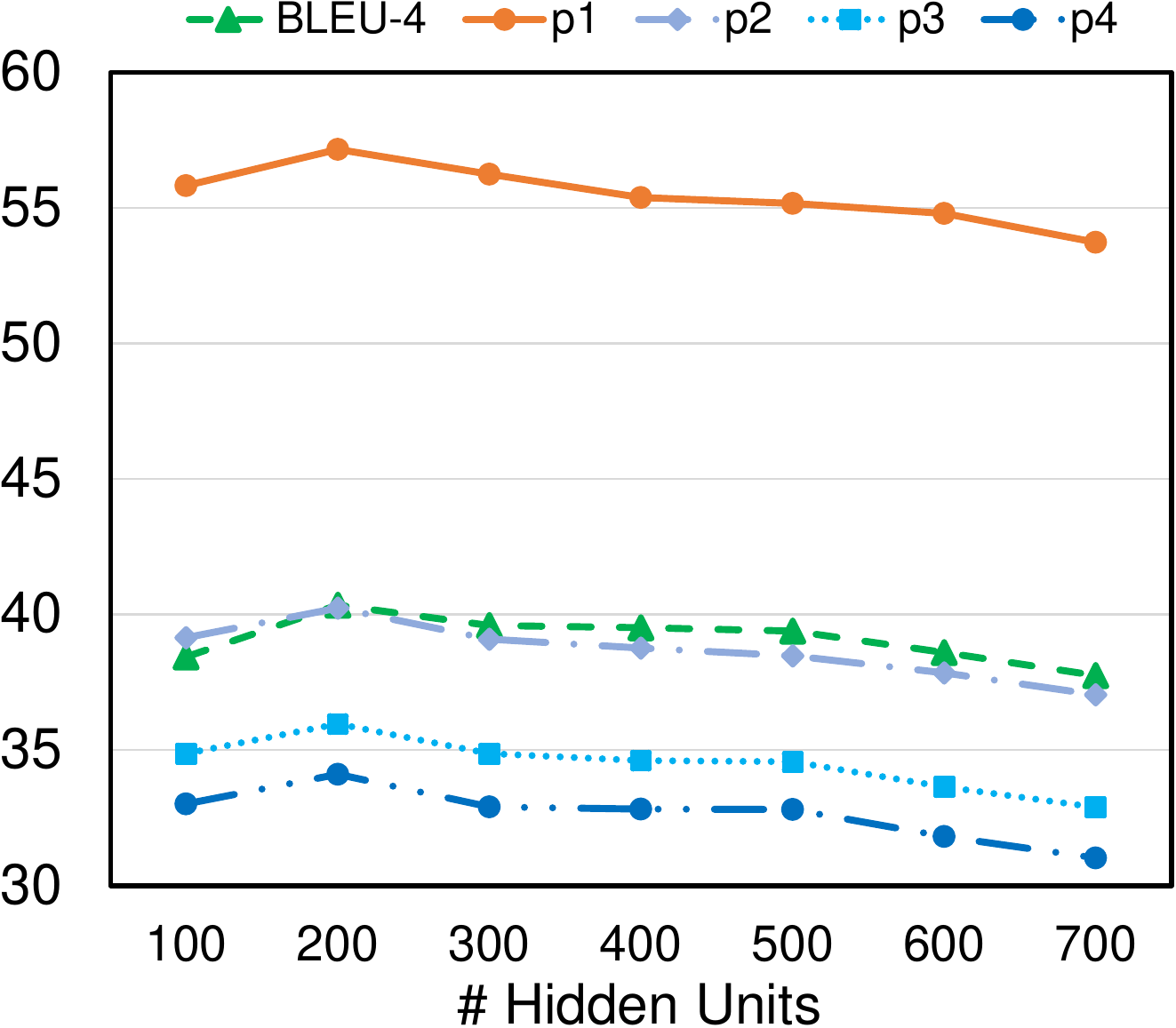}
        \caption{\small Different numbers of hidden units.}
        \end{subfigure}
        \begin{subfigure}[h]{0.45\textwidth}
        \centering
        \includegraphics[width=1 \textwidth]{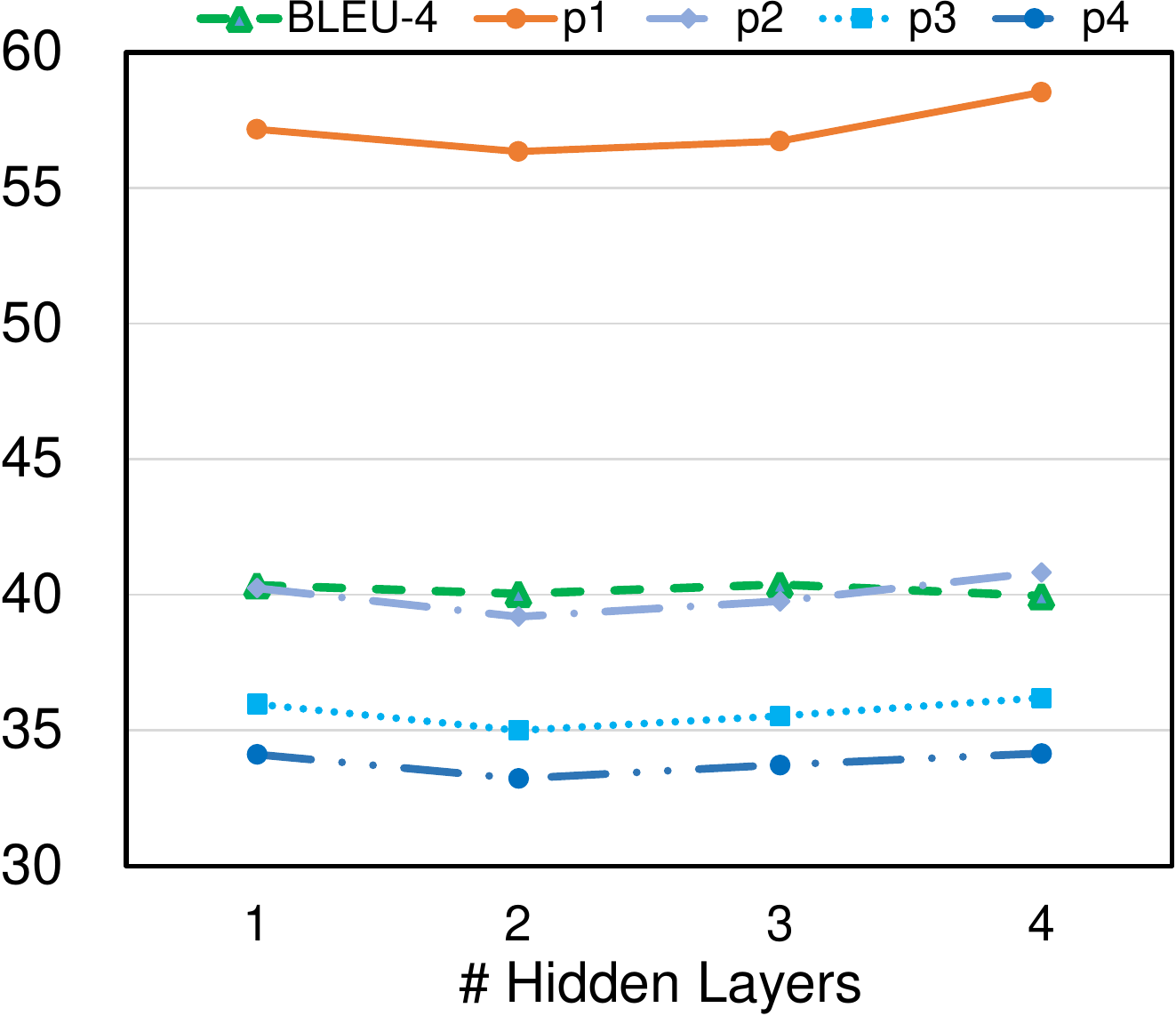}
        \caption{\small Different numbers of hidden layers.}
        \end{subfigure}
        \hfill
        \begin{subfigure}[h]{0.45\textwidth}
        \centering
        \includegraphics[width=1 \textwidth]{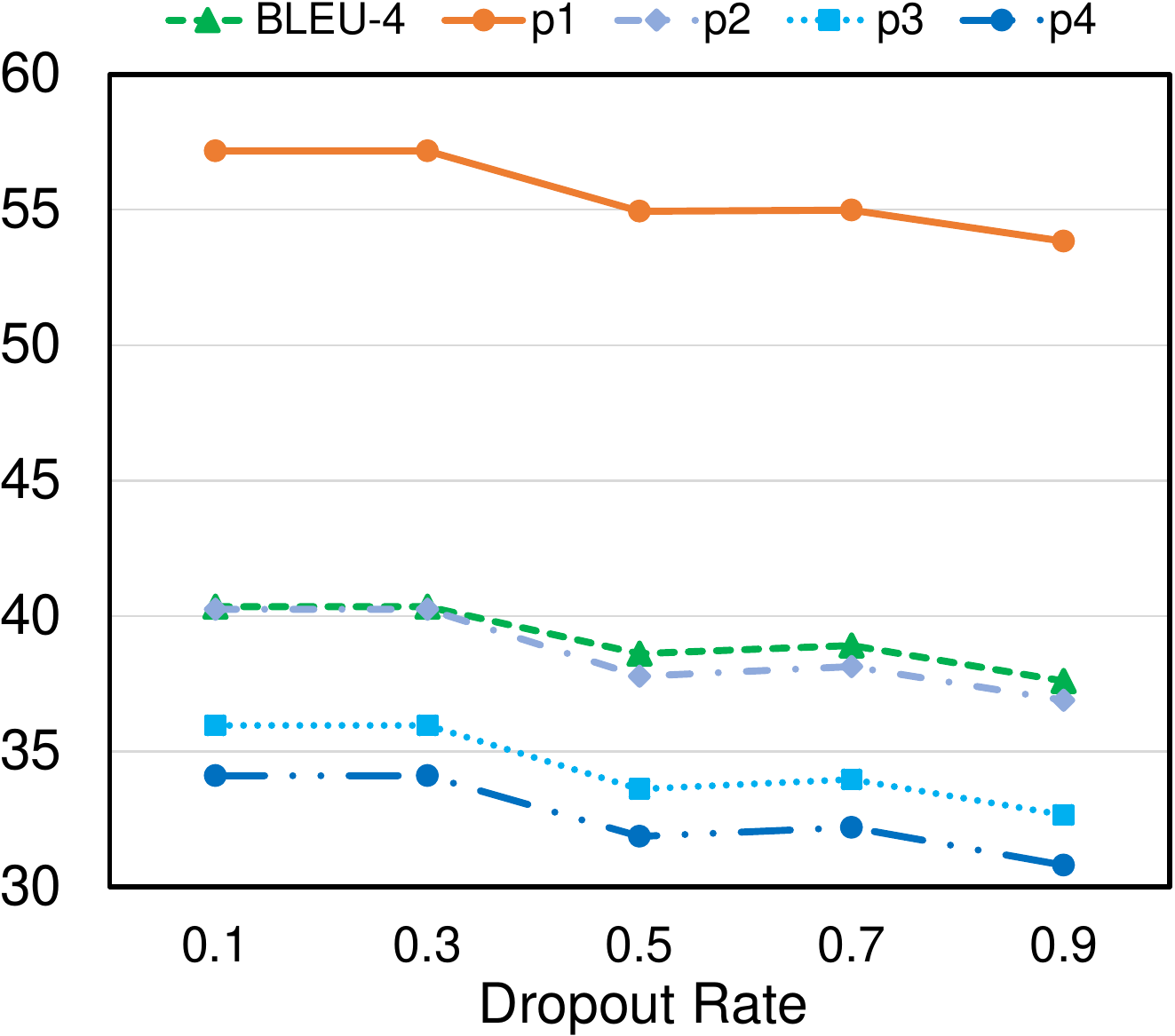}
        \caption{\small Different dropout rates.}
        \end{subfigure}
         \caption{Model performance under different parameter settings.}
     \label{fig:parameters}
\end{figure}

We also analyze the impact of different parameter settings on the model performance. Specifically, we compare the accuracy of \tool under varied parameters, including the number of retrieved responses, the number of hidden units, the number of hidden layers, dropout rate, and the dimension of word embeddings. Figure~\ref{fig:parameters} and Table~\ref{tab:embedding_size} show the influence of different parameter settings on the model performance. We observe that the accuracy of the model varies as the parameters change.

\begin{table}[t]
\centering
\caption{Impact of different dimensions of word embeddings on the performance of \tool.}\label{tab:embedding_size}
    \scalebox{1.0}{
        \begin{tabular}{c|r|r|r|r|r}
        \hline
        \hline
           \begin{tabular}[x]{@{}c@{}}Dimension\\of Word Embedding\end{tabular}   & BLEU-4 & $p_1$ & $p_2$ & $p_3$ & $p_4$\\
        \hline
        25 & 40.15 & 56.87 & 39.45 & 35.16 & 33.34 \\
        50 & 38.04 & 54.13 & 37.30 & 33.12 & 31.31 \\
        100 & \textbf{40.34} & \textbf{57.17} & \textbf{40.24} & \textbf{35.96} & \textbf{34.11}  \\
        200 & 39.09 & 55.20 & 38.42 & 34.13 & 32.27 \\ 
        \hline
        \hline
        \end{tabular}
    }
\end{table}

\textbf{\# Retrieved Responses:} As can be seen in Figure~\ref{fig:parameters} (a), with the number of retrieved responses increasing from 1 to 5, the BLEU-4 score fluctuates slightly, and when the number of retrieved responses is set as 4, \tool achieves the best performance. This indicates that more retrieved responses could be helpful for generating more accurate responses. However, since the relevance between the retrieved response and the review reduces as the number of retrieved responses increases, considering too many responses may bring interference to the final output.

\textbf{\# Hidden Units:} As shown in Figure~\ref{fig:parameters} (b), more hidden units may not be beneficial for improving accuracy. When the number of hidden units is larger than 200, the model performance exhibits a downward trend. Thus, we define the number of hidden units as 200 during the evaluation.

\textbf{\# Hidden Layers:} Figure~\ref{fig:parameters} (c) depicts the variations of the model performance as the number of hidden layers increase. We can observe that the variations are not obvious, ranging from 39.96 to 40.34 in terms of BLEU-4 score. Since with more hidden layers, both model training and testing time will increase, we set the number of hidden layers as 1 during the evaluation.

\textbf{Dropout Rate:} As can be seen in Figure~\ref{fig:parameters} (d), as the dropout rate grows, the model accuracy presents a decline trend, which implies that large dropout rates could greatly reduce the knowledge learnt by the previous layer, leading to poor generation performance. To reduce the information loss during the forward and backward propagation and avoid overfitting, the dropout rate is set as 0.1.

\textbf{Dimension of Word Embedding:} We compare the model performance under the four different dimensions of word embeddings provided by GloVe~\cite{glove} and the results are illustrated in Table~\ref{tab:embedding_size}. As can be seen, \tool achieves the poorest accuracy when the dimension of word embedding equals to 50 and the best when defined as 100. The performance decreases as the embedding dimension increases to 200, which indicates that more dimension may not be useful for enhancing the accuracy of the response generation. In this work, we set the dimension of word embeddings as 100.

\section{Human Evaluation}\label{sec:human}
In this section, we conduct human evaluation to further validate the effectiveness of the proposed \tool. The human evaluation is conducted through online questionnaire. We invite 20 participants totally, including 15 postgraduate students, four bachelors and one senior researcher, all of whom are not co-authors and major in computer science. Among the participants, 12 of them have industrial experience in software development for at least a year. Each participant is invited to read 25 user reviews and judge the quality of the responses generated by \tool, RRGen, and the official app developers. Each of them will be paid 10 USD if completing the questionnaire.

\subsection{Survey Design}
We randomly selected 100 review-response pairs and split them evenly into four groups, where each group consists of 25 review-response pairs. We create an online questionnaire for each group and ensure that each group is assessed by five different participants. In the questionnaire, each question describes one review-response pair, comprising one piece of user review, the developers' response, and its responses generated by RRGen and \tool. The order of the responses are randomly disrupted for each review.

\begin{figure}[t]
     \centering
    	\includegraphics[width=1 \textwidth]{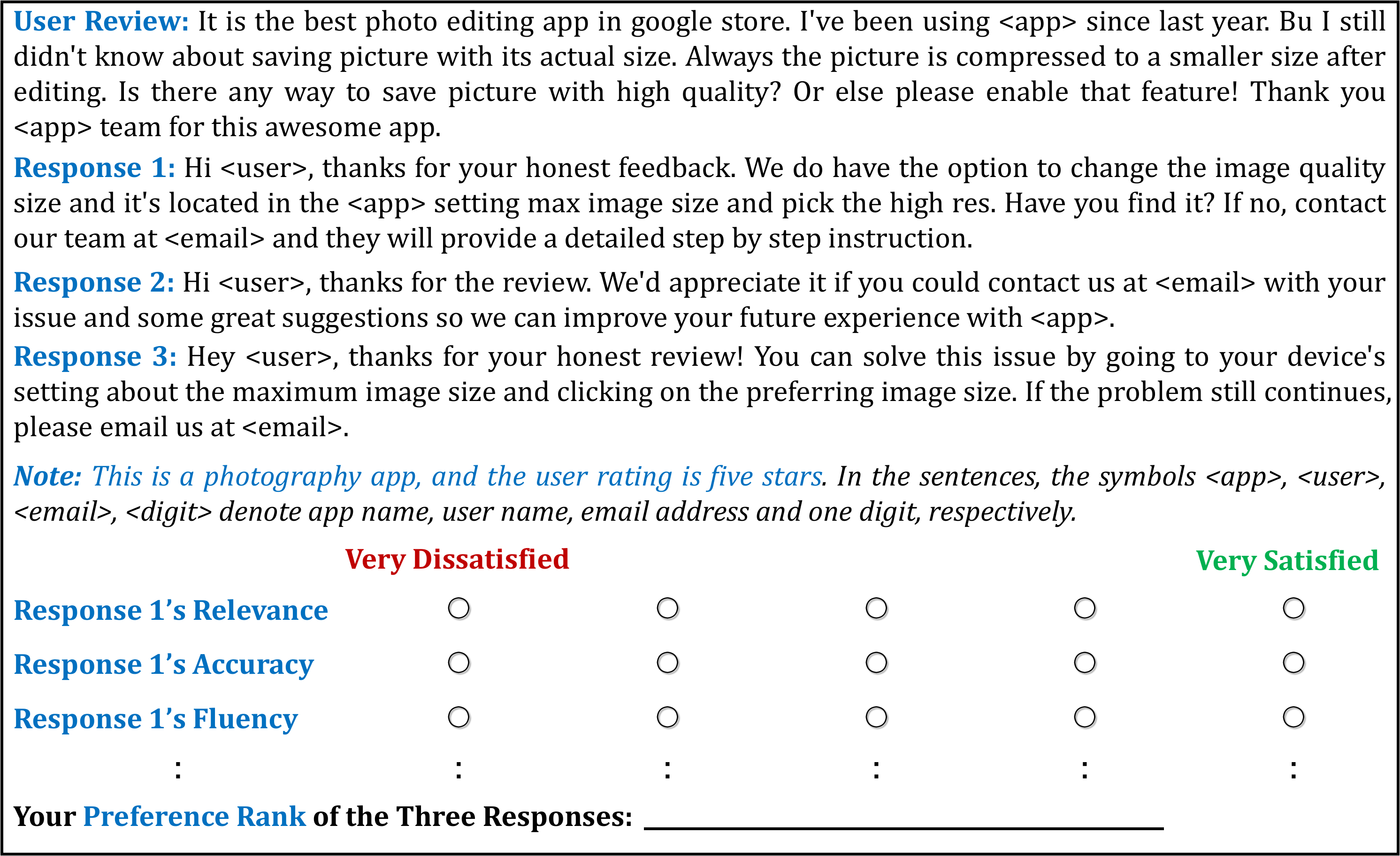}
         \caption{An example of questions in our questionnaires. Response 1, 2 and 3 correspond to the developer's response, the response produced by RRGen, and the output of \tool, respectively. The two-dot symbols indicate the simplified rating schemes for Response 2 and 3.}
     \label{fig:question_example}
\end{figure}

Following~\cite{DBLP:conf/kbse/GaoZX0LK19}, the quality of the responses is evaluated from three aspects, including ``\textit{grammatical fluency}'', ``\textit{relevance}'', and ``\textit{accuracy}''. We explained the three aspects at the beginning of each questionnaire: The metric ``\textit{grammatical fluency}'' measures the degree of the readability of the response; The metric ``\textit{relevance}'' estimates the extent of semantic relevance between the user review and response; And the metric ``\textit{accuracy}'' relates to the extent of the response accurately replying to the review. All the three aspects are scored based on 1-5 scale (1 for completely not satisfying the rating scheme and 5 for fully satisfying the rating scheme). Besides the three aspects, each participant is asked to rank the three responses based on the preference. The ``\textit{preference rank}'' score is evaluated on 1-3 scale (1 for the most preferred and 3 for the lease preferred). Figure~\ref{fig:question_example} shows one example of questions in our questionnaire. The participants are not aware of which response is written by developers or which one is generated by which model. They are asked to complete the online questionnaires separately. 

\subsection{Results}
\begin{figure}[t]
     \centering
    	\includegraphics[width=0.6 \textwidth]{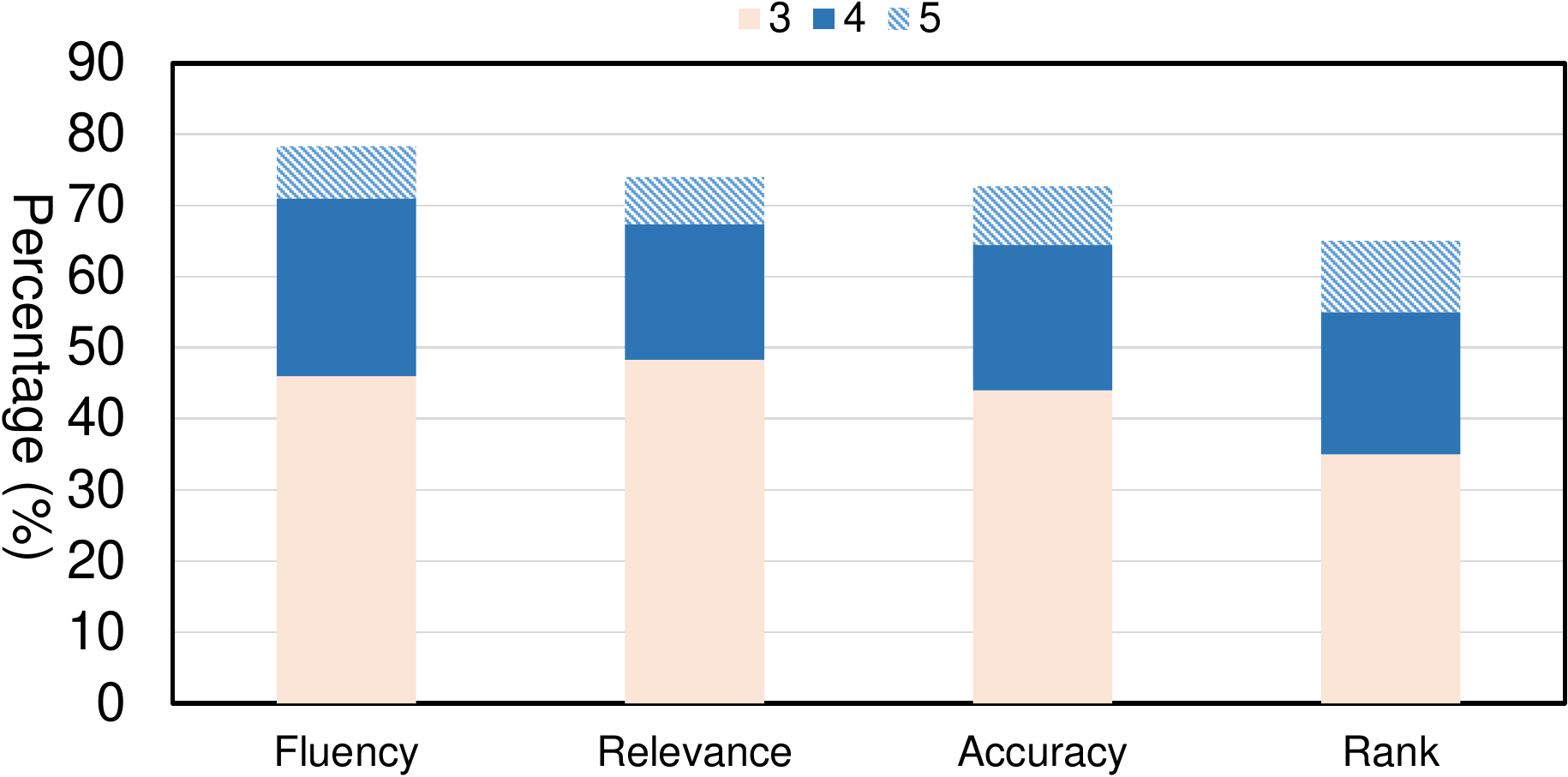}
         \caption{Agreement rate among the participants in the human evaluation. The horizontal axis and vertical axis indicate different evaluation metrics and the percentages of 3/4/5 participants giving the same scores, respectively.}
     \label{fig:agreement_rate}
\end{figure}

We finally received 500 sets of scores totally and five sets of scores for each review-response pair from the human evaluation. Each set contains scores regarding the four metrics, including ``\textit{grammatical fluency}'', ``\textit{relevance}'', ``\textit{accuracy}'' and ``\textit{preference rank}'', for the responses of \tool, RRGen, and official developers. The participants spent 1.72 hours on completing the questionnaire on average, with the median time cost at 1.40 hours. We compute the agreement rate on the four aspects given by the participants, illustrated in Figure~\ref{fig:agreement_rate}. As can be seen, 78.3\%, 74.0\%, 72.7\% and 65.0\% of the total 100 review-response pairs received at lease three identical scores regarding the ``\textit{grammatical fluency}'', ``\textit{relevance}'', ``\textit{accuracy}'' and ``\textit{preference rank}'' metrics respectively. Besides, 7.3\%, 6.7\%, 8.3\% and 10.0\% of the pairs are rated with consistent scores from the five annotators in terms of the respective metrics. This indicates that the participants achieved reasonable agreement on the quality of the generated responses. 

\begin{figure}[t]
     \centering
     \begin{subfigure}[h]{0.45\textwidth}
        \centering
    	\includegraphics[width=1 \textwidth]{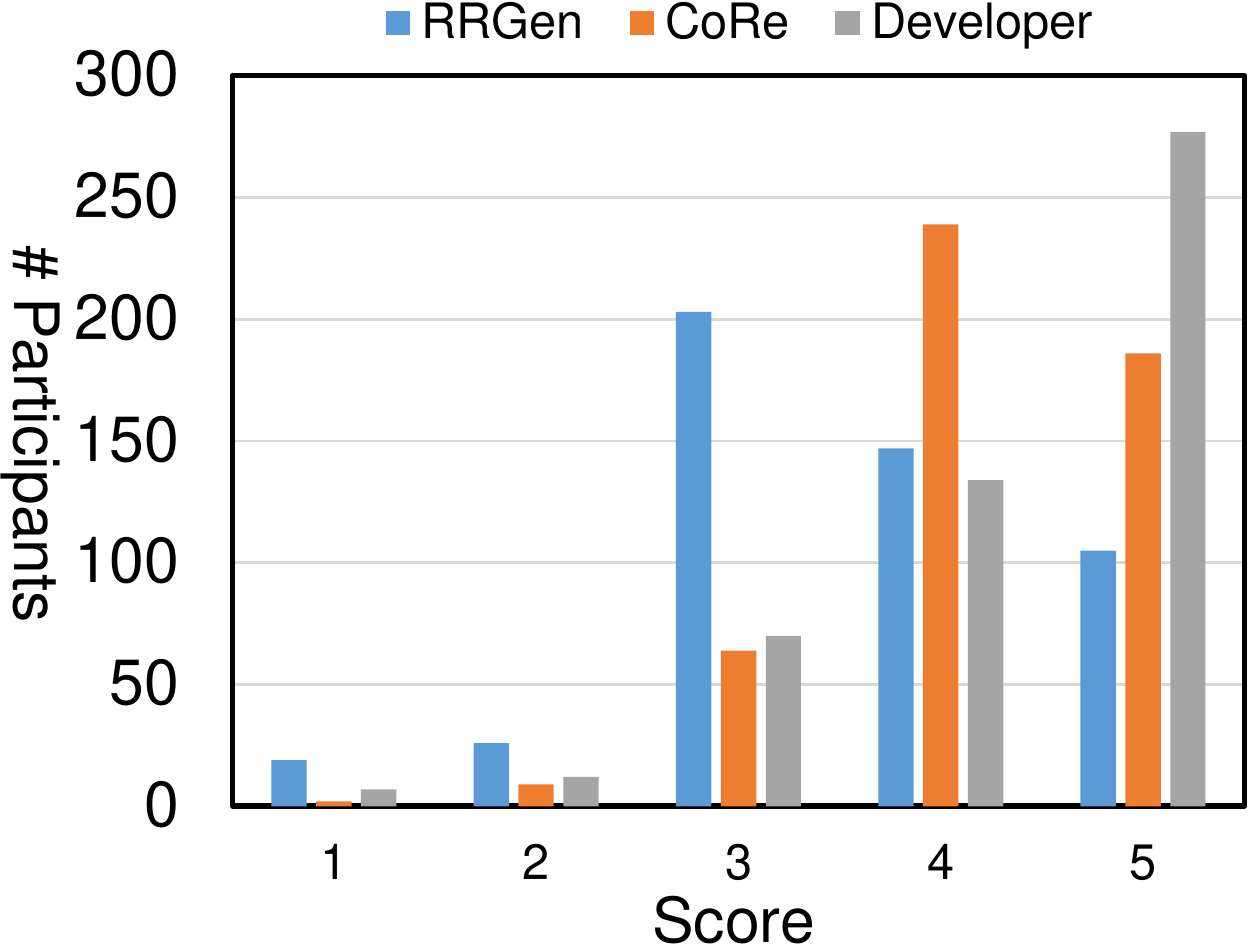}
    	\caption{\small Grammatical fluency.}
        \end{subfigure}
        \hfill
        \begin{subfigure}[h]{0.45\textwidth}
        \centering
        \includegraphics[width=1 \textwidth]{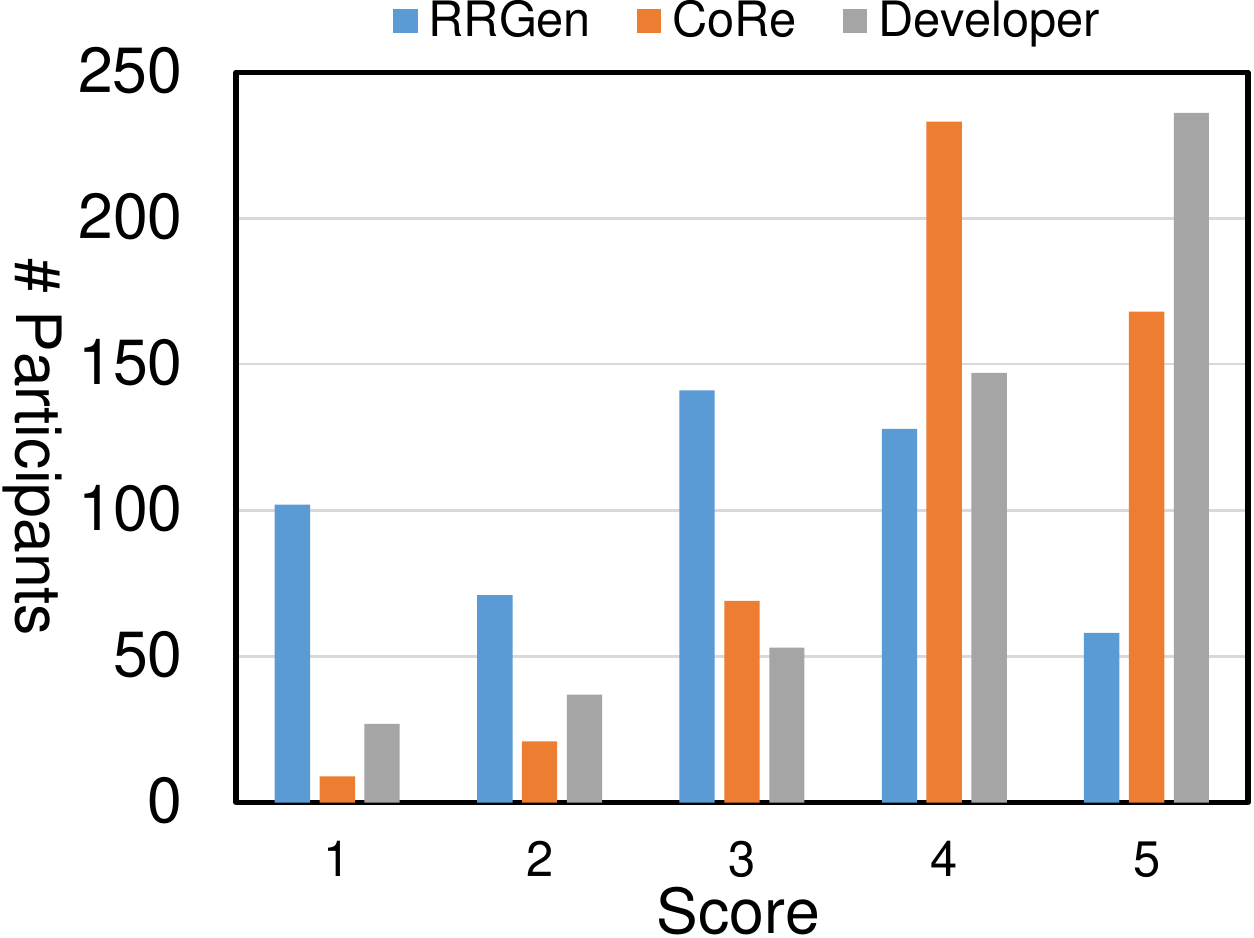}
        \caption{\small Relevance.}
        \end{subfigure}
        \begin{subfigure}[h]{0.45\textwidth}
        \centering
        \includegraphics[width=1 \textwidth]{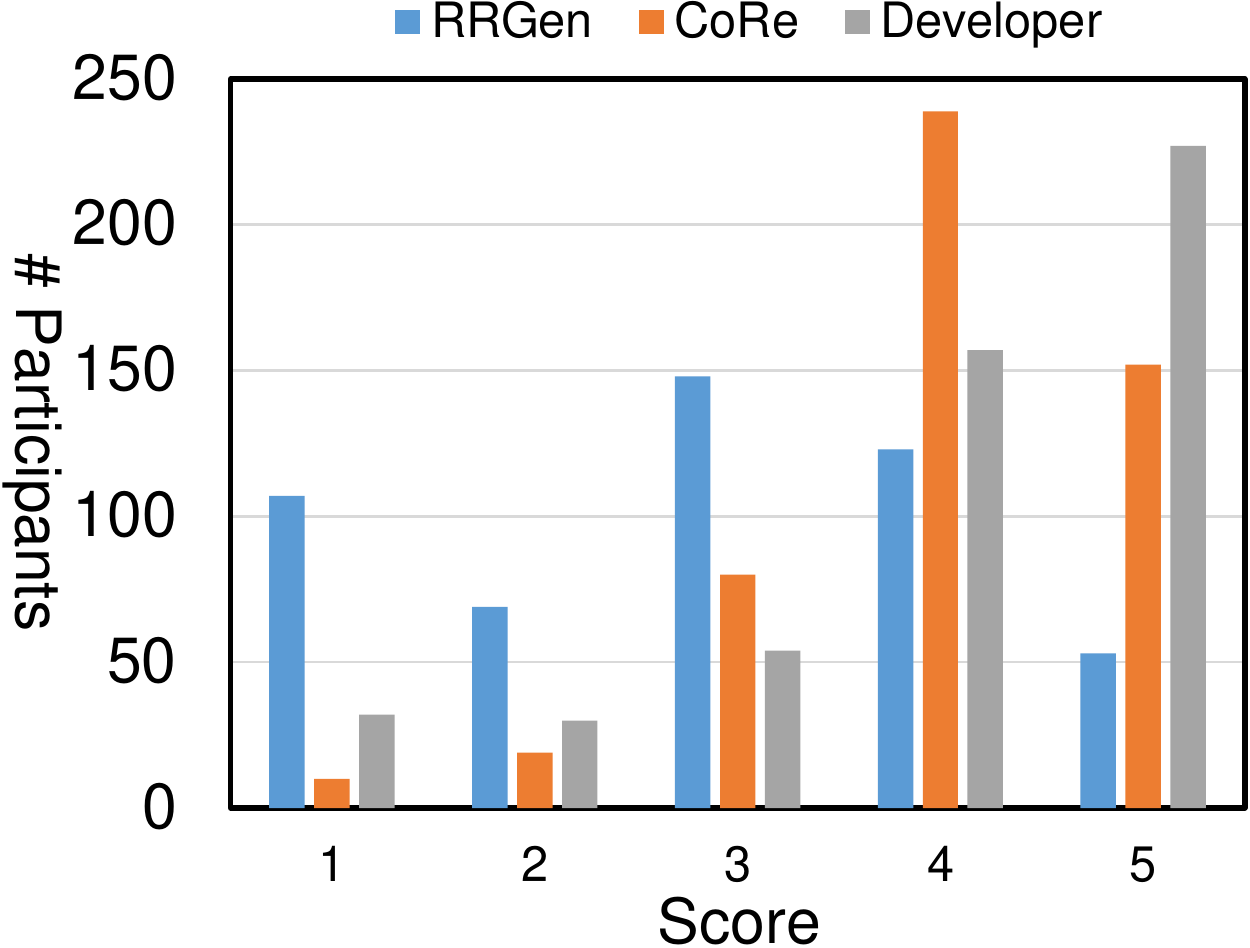}
        \caption{\small Accuracy.}
        \end{subfigure}
        \hfill
        \begin{subfigure}[h]{0.45\textwidth}
        \centering
        \includegraphics[width=1 \textwidth]{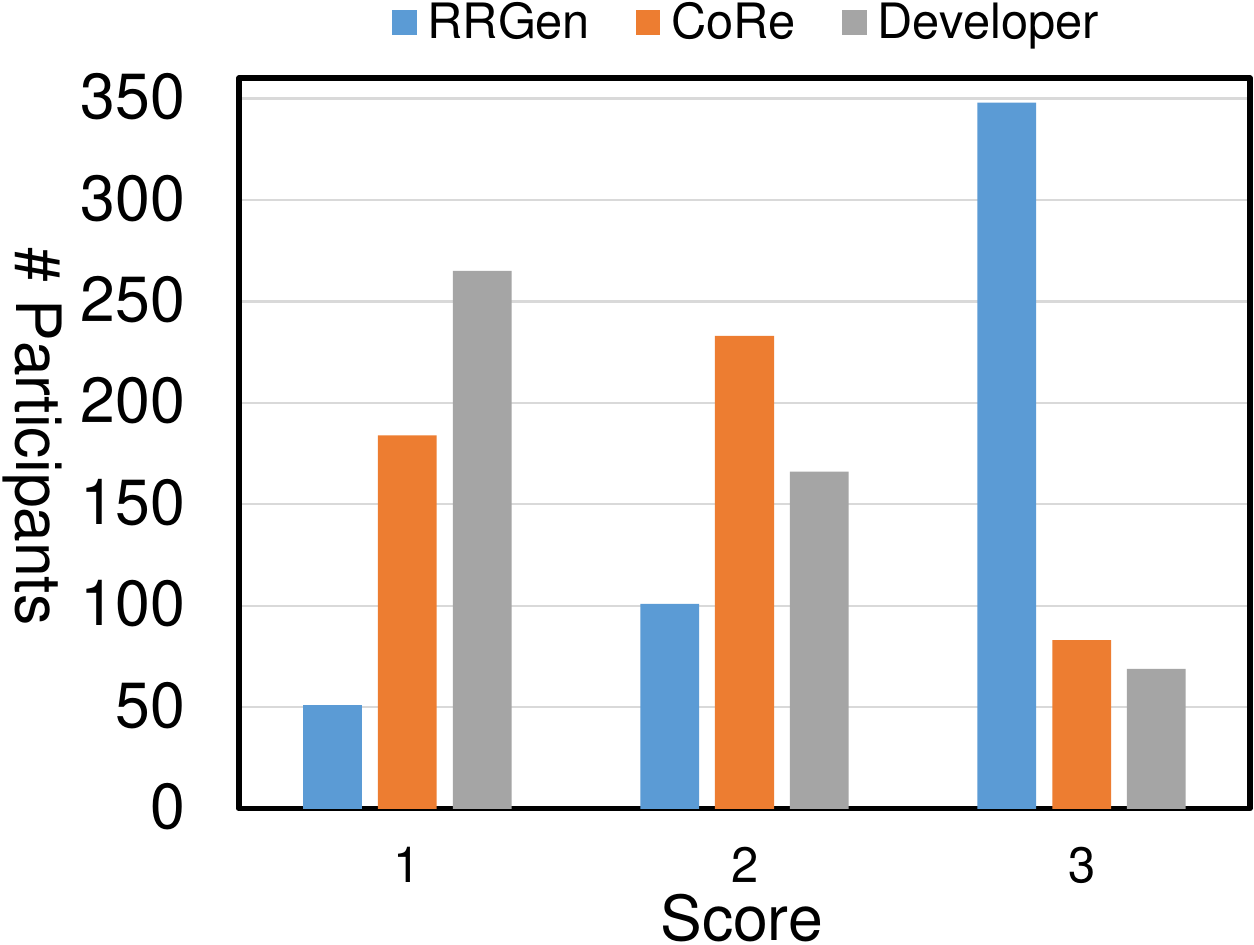}
        \caption{\small Preference rank.}
        \end{subfigure}
         \caption{Human evaluation results. For the metrics ``\textit{grammatical fluency}'', ``\textit{relevance}'' and ``\textit{accuracy}'', the higher scores the better; while for the metric ``\textit{preference rank}'', the lower scores the better. The vertical axis indicates the number of participants giving the scores.}
     \label{fig:human}
\end{figure}

Table~\ref{tab:human} and Figure~\ref{fig:human} depict the results of human evaluation. As can be seen, the responses from official developers receive the best scores from the participants among all the three responses and with respect to all the metrics. In terms of grammatical fluency, the average scores of the response generated by \tool and the developers' response are rather close, i.e., 4.19 and 4.32 respectively. As shown in Figure~\ref{fig:human} (a), most participants give the responses generated by RRGen a 3-star rating, while \tool receives more 4/5-star ratings. This indicates that \tool can produce more grammatically fluent responses than RRGen. Regarding the relevance, the responses generated by RRGen are rated much poorer than those output by \tool. Combined with Figure~\ref{fig:human} (b), we can observe that the more than half (62.5\%) of the participants enter ratings lower than 4 for the responses generated by RRGen, and the number of 4/5-star ratings for the responses produced by \tool is 1.15 times than those for the responses of RRGen. Developers' responses receive the most 5-star ratings comparing to the generated responses. This implies that the responses output by \tool tend to be more relevant to the reviews than those generated by RRGen. In terms of the ``\textit{accuracy}'' metric, we find that the average scores for the responses output by \tool and the developer's responses are much close, i.e., 4.00 and 4.03 respectively. As illustrated in Figure~\ref{fig:human} (c), the responses generated by \tool receive slightly more 4/5-star ratings than the developers' responses (391 v.s. 384), and 1.22 times than the responses generated by RRGen (176). The result demonstrate that \tool can produce accurate responses to the user reviews, which is also reflected in the distributions of the ``\textit{preference rank}'' scores, as shown in Figure~\ref{fig:human} (d). We can discover that most participants rank the responses output by RRGen as the least preferred (69.6\%) and the developers' responses as the most favored (53.0\%), and the responses of \tool present similar preference score as the developers' responses on average, i.e., 1.79 v.s. 1.60 (as shown in Table~\ref{tab:human}). The human study further validates the effectiveness of the proposed \tool for review response generation.

\begin{table}[t]
\centering
\caption{Comparison results based on human evaluation. Average scores are computed and \textbf{bold} indicates top scores. Two-tailed t-test results between \tool and RRGen are indicated with *($p$-value<0.01).}\label{tab:human}
    \scalebox{1.0}{
        \begin{tabular}{c|r|r|r|r}
        \hline
        \hline
          &  \begin{tabular}[x]{@{}c@{}}Grammatical\\Fluency\end{tabular}
           & Relevance & Accuracy & \begin{tabular}[x]{@{}c@{}}Preference\\Rank\end{tabular} \\
        \hline
        RRGen & 3.58* & 2.93* & 2.89* & 2.59* \\
        \tool & 4.19 & 4.06 & 4.00 & 1.79 \\
        \hline
        Developer & \textbf{4.32} & \textbf{4.56} & \textbf{4.03} & \textbf{1.60} \\
        \hline
        \hline
        \end{tabular}
    }
\end{table}

\section{Discussion}\label{sec:discussion}
In this section, we discuss the advantages, limitations, and threats of our model.

\subsection{Why does Our Model Work?}\label{sec:advantage}
We have conducted a deep analysis on the advantages of combining app descriptions and retrieved responses for review response generation in \tool.

\subsubsection{App descriptions} 
\begin{figure}[t]
     \centering
    	\includegraphics[width=0.9 \textwidth]{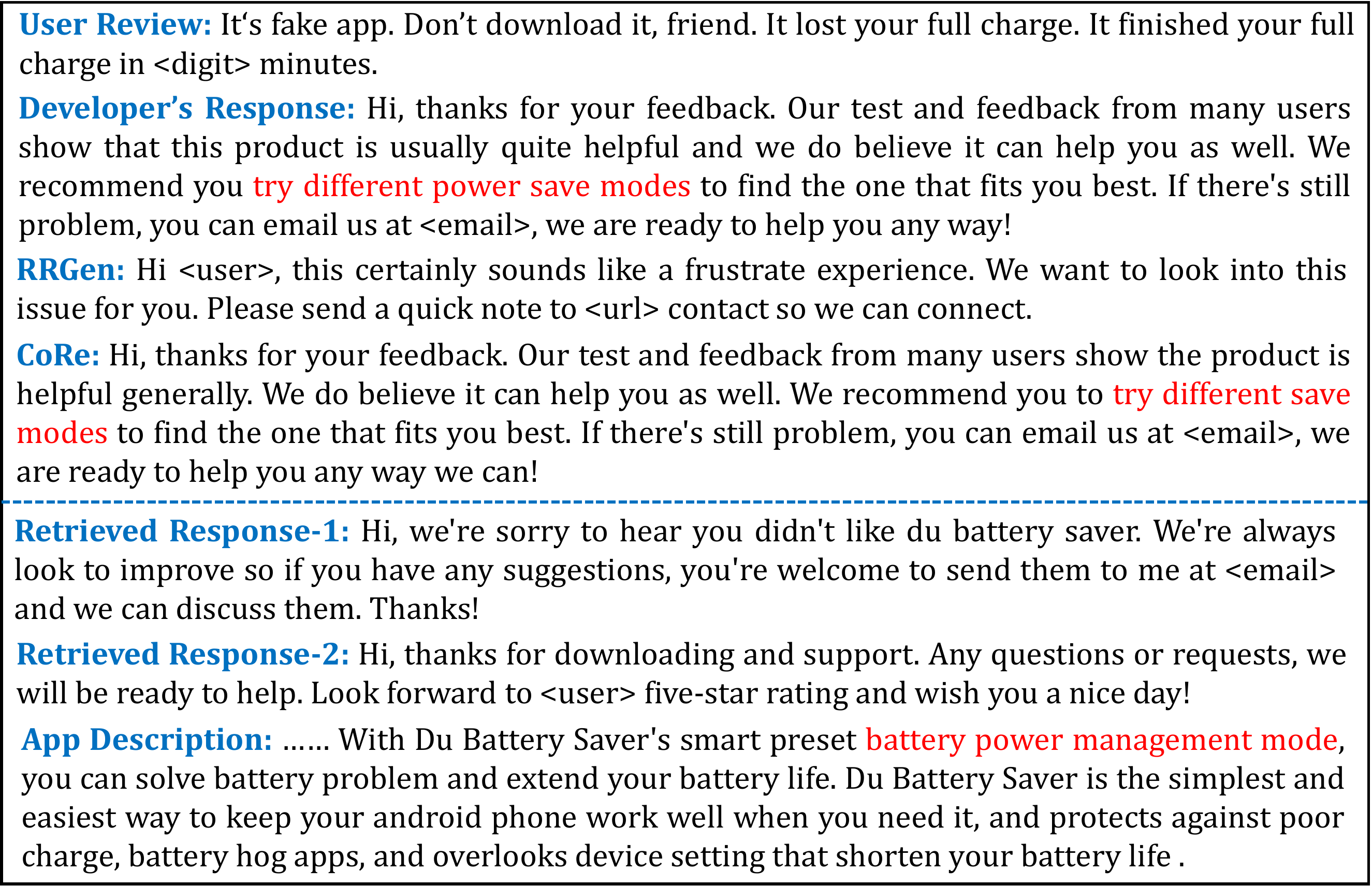}
         \caption{A user review with the generated response where \tool can generate responses based on the app description. The fonts in red are indicative of the partial topical words in corresponding texts. We only illustrate the responses of the top two retrieved reviews here for saving space.}
     \label{fig:description_example}
\end{figure}

\begin{figure}[t]
     \centering
    	\includegraphics[width=1 \textwidth]{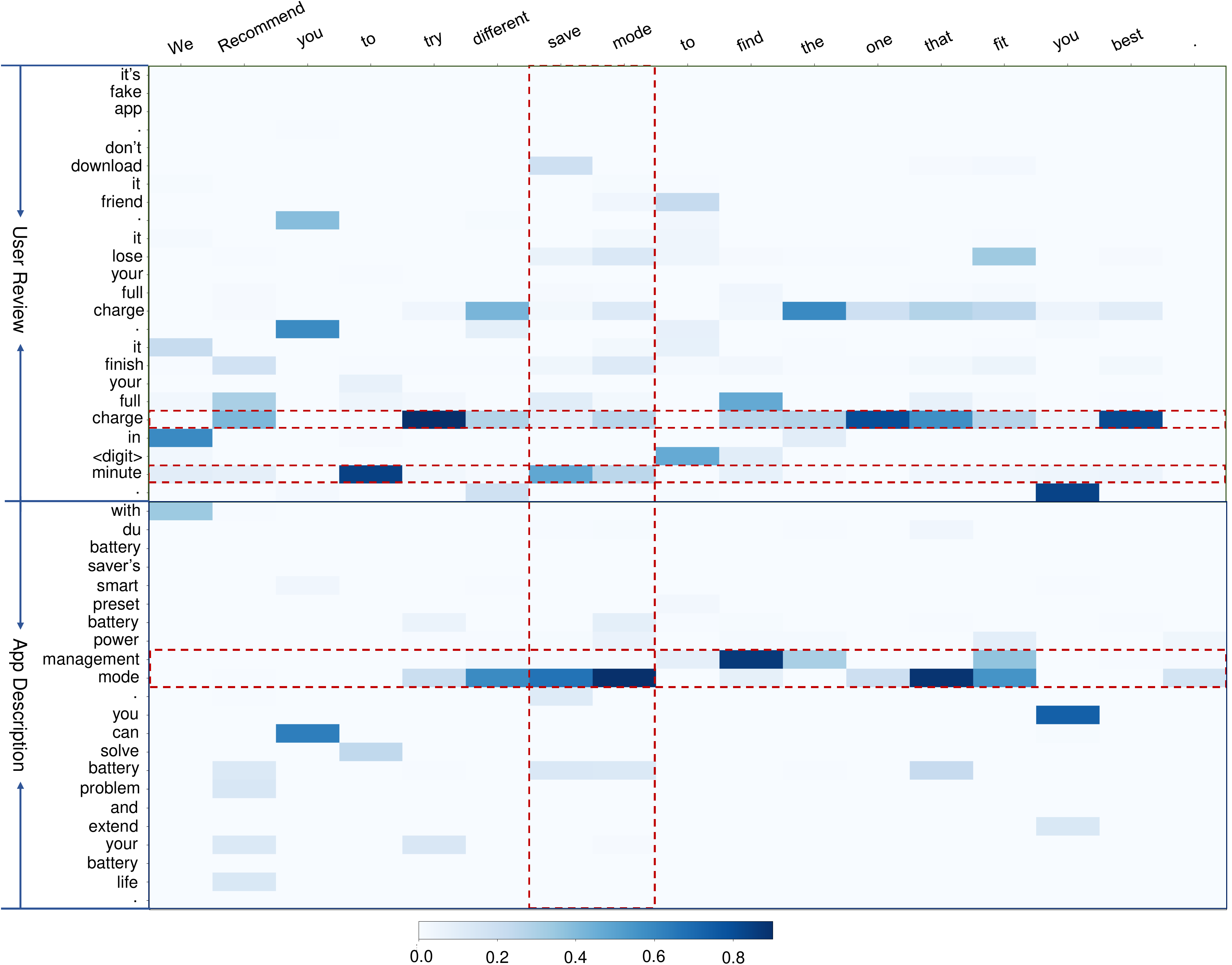}
         \caption{A heatmap representing the alignment between the user review (left-top)/app description (left-bottom) and generated response by \tool (top). The columns represent the distribution over the user review/app description after generating each word in the response. Darker colors indicate higher attention weights and manifest a stronger correlation between the target word and source word. Red dotted rectangles highlight partial topical words in corresponding texts.}
     \label{fig:vis_attention}
\end{figure}

App descriptions generally contain keywords related to main app features, aiming at convincing users to download the apps and facilitating user search through app stores. By considering app descriptions, \tool can recognize the topics/functionalities discussed by users more accurately. For example, it can learn that the review ``\textit{It lose your full charge.}'' is related to the ``\textit{power save mode}'' in the app, and generate response providing the solution ``\textit{trying different save mode}'', as shown in Figure~\ref{fig:description_example}; while the response generated by RRGen is rather in general purpose and not topically relevant to the review. Figure \ref{fig:vis_attention} visualizes the latent alignment over the user review/app description based on the attention weights $\alpha_{tj}$ from Equation~(\ref{equ:attention}) and $\alpha_t^{(r)}$ Equation~(\ref{equ:att_desc}) respectively. Each column indicates the word distribution over the user review/app description during response generation, which implies the importance of the words in the user review/app description when generating the target word in the response. We can observe the obvious correlation between the word ``\textit{mode}'' (in the app description) and ``\textit{save mode}'' (in the response), and relatively weak correlations between ``\textit{charge}''/``\textit{minute}'' (in the review) and ``\textit{save mode}'' (in the response). This illustrates that \tool can build implicit relations between the topical words in app descriptions and corresponding responses, which can help generate relevant and accurate response given a review.

\subsubsection{Retrieved responses} 
\begin{figure}[t]
     \centering
    	\includegraphics[width=0.9 \textwidth]{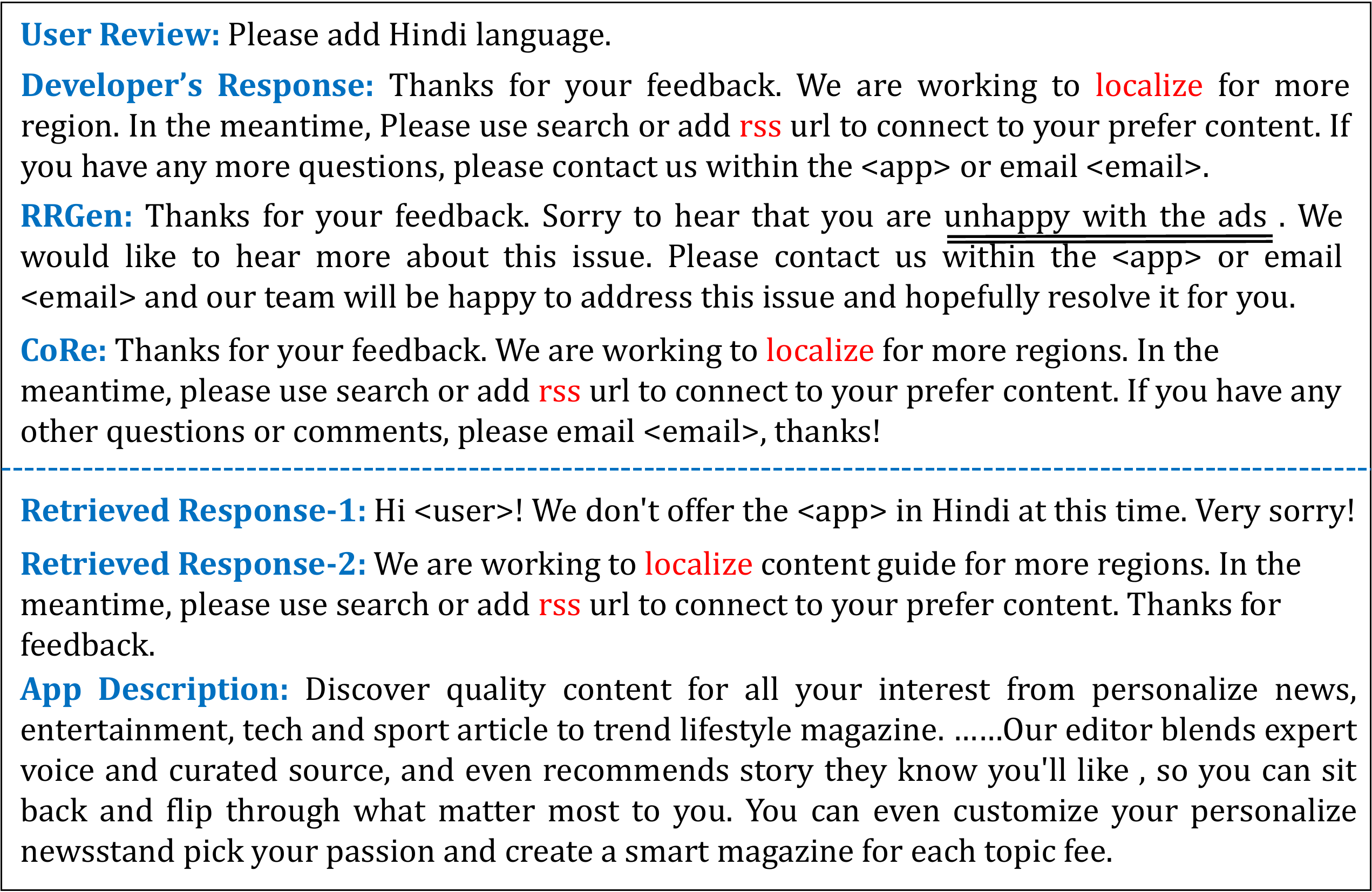}
         \caption{A user review with the generated response where \tool can generate responses with low-frequency words. The fonts in red are indicative of the low-frequency words (frequency$\leq$100) and the double-underlined words mean they are topically irrelevant to the user review. Responses of the retrieved top two reviews and the app description are also illustrated.}
     \label{fig:example_low_freq}
\end{figure}

NMT-based approaches tend to prefer high-frequency words in the corpus, and the generated responses are often generic and not informative~\cite{DBLP:conf/emnlp/ArthurNN16,DBLP:conf/naacl/ZhangUSNN18,DBLP:conf/cikm/0005HQQGCLSL19}. For example, they may fail for the responses containing low-frequency words. In our experiment, we find that 51,364/309,246 (16.61\%) responses in the corpus contain low-frequency words (frequency$\leq$100). Since similar reviews based on IR-based methods are generally related to the same semantics, their responses could be semantically related and the words in the expected responses (including the low-frequency ones) are also highly probable to appear in them. For example, for the review in Figure~\ref{fig:example_low_freq}, we retrieve most similar reviews with respect the semantics (i.e., tf-idf representations in the paper) from the training corpus. We only present the responses of the top two similar reviews here for saving space. We can see that the low-frequency words ``\textit{localize}'' and ``\textit{rss}'' (which is an abbreviation of Really Simple Syndication, a web feed that allows users to access updates of websites in a standardized format) also appears in the retrieved response (i.e., Retrieved response-2). The words are ignored by RRGen but correctly predicted by \tool since they appear in the retrieved responses and are effectively captured during attention fusion (Section~\ref{sec:fusion}). In contrast, the response generated by RRGen is topically irrelevant to the review, supposing the review is talking about ``\textit{ads}''. This exhibits that the retrieved responses in \tool are helpful for generating the responses with low-frequency words.


\subsection{Limitations of \tool}
Although the proposed \tool enhances the performance of review reply generation, \tool does not handle two case types well, including the reviews that do not require responses and the reviews with poor responses generated by \tool. For the first  case type, we refer readers to the work~\cite{DBLP:conf/ease/SrisophaSLB20,DBLP:journals/ese/HassanTBH18} on summarizing which review features spur the responses. In this work, we are more focused on the subsequent behavior for developers, i.e., responding to the reviews requiring responses. For the second case type, we design a quality assurance filter based on the manually-annotated review-response pairs in Section~\ref{sec:human} to automatically learn the cases in which the proposed \tool does not perform well. The poorly-generated responses can be delegated to developers for further inspection before posting.

\textbf{Filter Design:} The proposed quality assurance filter contains three main steps. We first prepare the gold set for filter training. We employ the involved reviews and the corresponding responses generated by \tool in the human evaluation as our gold set. Each review and the corresponding generated message are associated with scores which indicate the extent of accuracy to reply to the review (as shown in Figure \ref{fig:human}). To be conservative, we labeled the reviews that receive the ``\textit{accuracy}'' score of one, two or three from one annotator as ``bad'' and all the other reviews as ``not bad''. Then we extract the unigram tf-idf representations of the reviews as the features, since tf-idf has been widely used in natural language processing for feature representation~\cite{DBLP:journals/tse/FanXLH20,DBLP:conf/kbse/VuNPN15}. We finally train a Gaussian kernel SVM using stochastic gradient descent (SGD) as the learning algorithm based on the dataset of reviews and their labels. The trained SVM will be adopted to predict whether the \tool model generates a ``bad'' response for a user review.

\textbf{Filter Performance:} We split gold set into 10 folds based on stratified shuffle. For each fold, we train a SVM model on the other 9 folds, and test the SVM model on the one fold. We finally obtained the test results for every fold. Table~\ref{tab:filter} shows the predicts of all the folds. In terms of detecting reviews for which the \tool model will generate ``bad'' responses, the filter has 83.0\% precision and 93.6\% recall. Furthermore, it can reduce 31.8\% of the ``bad'' responses. The results demonstrate the usefulness of the proposed filter component for detecting the poorly-generated responses. We also deployed the trained filter to the test set used in Section~\ref{sec:result} and observed that the model performance showed 40.55 in terms of BLEU-4 score with 2,106/14,727 (14.3\%) ``bad'' responses removed, which is slightly higher than the BLEU-4 score (40.34) reported in our earlier experiment using all the test samples. Developers can focus on examining the ``bad'' responses during using the proposed \tool model. For the other reviews, developers can directly adopt the responses generated by \tool. 

\begin{table}[t]
\centering
\caption{Predicted results of the cross evaluation of the quality assurance filter.}\label{tab:filter}
    \scalebox{1.0}{
        \begin{tabular}{c|c| r|r}
        \hline
        \hline
       \multicolumn{2}{c|}{\multirow{2}{*}{Predicted Results}} & \multicolumn{2}{c}{Actual Labels} \\
       \cline{3-4}
       \multicolumn{2}{c|}{}& Not Bad &  Bad \\
         \hline
        \multirow{2}{*}{\begin{tabular}[x]{@{}c@{}}Predicted\\Labels\end{tabular}} & Not Bad & 73 & 15 \\
        & Bad & 5 & 7 \\
        \hline
        \hline
        \end{tabular}
    }
\end{table}

\subsection{Threats to Validity}
There are three main threats to the validity of our study.
\begin{enumerate}
   \item The scale of dataset. We directly use the publicly released data of RRGen provided by their authors. The data include only review-response pairs of 58 free apps from Google Play Store. The limited categories and number of studied apps may influence the generalization of the proposed \tool. Since the dataset is the only one with huge quantities of review-response pairs at this time, we will eliminate this threat as soon as larger-scale datasets are publicly available.
    \item The retrieved reviews may not always present high similarities. One of the reasons may be the similarity measurement approach is simply based on tf-idf representations, in which the tf-idf may not be the best approach to represent the semantics of the review texts~\cite{DBLP:conf/nips/MikolovSCCD13}. Another reason is the available review-response pairs may be limited. Since involving more complex approach for retrieving similar reviews could increase the burden of model training and the effectiveness of tf-idf in review representation has already been demonstrated in~\cite{DBLP:conf/kbse/VuNPN15}, we investigate the light-weight tf-idf approach in the paper. We will explore the impact of different retrieval approaches and datasets on automatic review response generation in the future.
    \item Bias in manual inspection. The results of the human evaluation can be impacted by the participants' experience and their understanding of the evaluation metrics. To mitigate the bias in manual inspection, we ensure that each review-response pair was evaluated by five different participants. Besides, we randomly disrupt the order of the three types of responses for each review, so that the influence of participants' prior knowledge about the response orders is eliminated.
\end{enumerate}

\section{Related Work}\label{sec:literature}
We split the related work into three categories: 1) the work that conducts app review mining; 2) the work that analyzes user developer dialogue; and 3) the work that generates short text conversational.

\subsection{App Review Mining}
App reviews are a valuable resource provided directly by the customers, which can be exploited by app developers during the bug-fixing~\cite{DBLP:journals/tse/AliGA13} and feature-improving process~\cite{DBLP:conf/issre/GaoWHZZL15}. The essence of app review mining lies in the effective extraction and summarization of the useful information from app reviews. Iacob et al.~\cite{DBLP:conf/bcshci/IacobVH13} manually label 3,278 reviews of 161 apps into nine classes, and discover that 23.3\% of the feedback constitutes requirements from users, e.g., various issues encountered by users. Due to the ever-increasing amount of reviews, previous studies resort to generic NLP techniques to automate the information extraction process. For example, Iacob and Harrison~\cite{DBLP:conf/msr/IacobH13} use pre-defined linguistic rules for retrieving feature requests from app reviews. Di Sorbo et al.~\cite{DBLP:conf/sigsoft/SorboPASVCG16} build a two-level classifiers to summarize the enormous amount of information in user reviews, where user intentions and review topics are respectively classified. Developers can learn feature requests and bug reports more quickly when presented with the summary. \cite{DBLP:conf/icse/VillarroelBROP16}, \cite{DBLP:conf/icse/ChenLHXZ14}, \cite{DBLP:conf/issre/GaoWHZZL15}, and \cite{DBLP:conf/kbse/VuNPN15}, etc., employ unsupervised clustering methods to prioritize user reviews for better app release planning. Nayebi, Farrahi, and Ruhe~\cite{DBLP:conf/esem/NayebiFR17} adopt app reviews besides other release attributes for predicting release marketability and determining which versions to be released.

Another line of work on app review mining is about predicting user sentiment towards the app features or functionalities~\cite{DBLP:conf/re/GuzmanM14,DBLP:conf/kbse/GuK15,DBLP:conf/www/LuizVAMSCGR18,DBLP:journals/jss/Genc-NayebiA17}. For example, Guzman et al.~\cite{DBLP:conf/re/GuzmanM14} use topic modeling techniques to group fine-grained features into more meaningful high-level features and then predict the sentiment associated with each feature. Instead of treating reviews as bags-of-words (i.e., mixed review categories), Gu and Kim~\cite{DBLP:conf/kbse/GuK15} only consider the reviews related to aspect evaluation and then estimate the aspect sentiment based on a pattern-based parser.

\subsection{Analysis of User Developer Dialogue}
Analysis of user developer dialogue explores the rich interplay between app customers and their developers \cite{DBLP:journals/infsof/FinkelsteinHJMS17}. Oh et al. et al.~\cite{DBLP:conf/chi/OhKLLS13} discover that users tend to take a passive action such as uninstalling apps when their inquires (e.g., user reviews) would take long time to be responded or receive no response. Srisopha et al.~\cite{DBLP:conf/ease/SrisophaSLB20} investigate which features of user reviews spur developers' responses, and find that ratings, review length and the proportions of positive and negative words are the most important features to predict developer responses. Both McIlroy et al.~\cite{DBLP:journals/software/McIlroySAH17} and Hassan et al.~\cite{DBLP:journals/ese/HassanTBH18}'s studies observe the positive impact of developers' responses on user ratings, for example, users would change their ratings 38.7\% of the time following a response. To alleviate the burden in the responding process, Gao et al.~\cite{DBLP:conf/kbse/GaoZX0LK19} propose an NMT-based approach named RRGen for automatically generating the review responses.

\subsection{Short Text Conversation Generation}
Short text conversation is one of the most challenging natural language processing problems, involving language understanding and utilization of common sense knowledge~\cite{DBLP:conf/acl/ShangLL15}. Short text conversation can be formulated as a ranking or a generation problem. The former formulation aims at learning the semantic matching relations between conversation histories and responses in the knowledge base, and retrieving the most relevant responses from the base for the current conversation. Ranking-based approaches have the advantage of returning fluent and informative responses, but may fail to return any appropriate responses for those unseen conversations. The generation-based formulation treats generation of conversational dialogue as a data-driven statistical machine translation (SMT)~\cite{DBLP:conf/emnlp/RitterCD11,DBLP:conf/flairs/ChenTALT11}, and has been boosted by the success of deep learning models~\cite{DBLP:conf/nips/SutskeverVL14} and reinforcement learning approaches~\cite{DBLP:conf/emnlp/LiMRJGG16}. Gao et al.~\cite{DBLP:journals/ftir/GaoGL19} perform a comprehensive survey of neural conversation models in this area. The major problem of the generation-based approaches is that the generated responses are often generic and not informative due to the lack of grounding knowledge~\cite{DBLP:conf/cikm/0005HQQGCLSL19}. In this work, we propose to integrate contextual knowledge, including app descriptions and retrieved responses, for accurate review response generation.

\section{Conclusions and Future Work}\label{sec:conclusion}
This paper proposes \tool, a novel framework aiming at automatically generating accurate responses for user reviews and thereby ensuring a good user experience of the mobile applications. We present that employing app descriptions and the responses of similar user reviews in the training corpus as contextual knowledge is beneficial for generating high-quality responses. Both quantitative evaluation and human evaluation show that the proposed model \tool significantly outperforms the baseline models. The encouraging experimental results demonstrate the importance of involving contextual knowledge for accurate review response generation. We also analyze the advantages and limitations in this work, and plan to address them in the future.



\bibliographystyle{ACM-Reference-Format}
\bibliography{sigproc}


\end{document}